\documentclass[pdflatex,sn-mathphys-num]{sn-jnl}

\usepackage{graphicx}%
\usepackage{multirow}%
\usepackage{amsmath,amssymb,amsfonts}%
\usepackage{amsthm}%
\usepackage{mathrsfs}%
\usepackage[title]{appendix}%
\usepackage{xcolor}%
\usepackage{textcomp}%
\usepackage{manyfoot}%
\usepackage{booktabs}%
\usepackage{tabularx}
\usepackage{algorithm}%
\usepackage{algorithmicx}%
\usepackage{algpseudocode}%
\usepackage{listings}%
\usepackage[normalem]{ulem}
\lstdefinestyle{jsonstyle}{
    basicstyle=\ttfamily\small,
    breaklines=true,
    frame=single,
    columns=fullflexible,
    showstringspaces=false,
    tabsize=2,
    captionpos=b
}

\definecolor{mdgray}{RGB}{245,245,245}
\definecolor{mdframe}{RGB}{210,210,210}
\definecolor{mdcomment}{RGB}{100,100,100}
\definecolor{mdstring}{RGB}{40,90,150}
\lstdefinestyle{markdownstyle}{
    basicstyle=\ttfamily\small,
    backgroundcolor=\color{mdgray},
    frame=single,
    rulecolor=\color{mdframe},
    breaklines=true,
    breakatwhitespace=false,
    columns=fullflexible,
    keepspaces=true,
    showstringspaces=false,
    tabsize=2,
    captionpos=b,
    upquote=true,
    commentstyle=\color{mdcomment},
    stringstyle=\color{mdstring}
}

\theoremstyle{thmstyleone}%
\theoremstyle{thmstyletwo}%

\theoremstyle{thmstylethree}%

\begin{document}
\title[SmartIntent: A Serverless Edge Intelligence Architecture with Local LLMs for Building Automation]{SmartIntent: A Serverless Edge Intelligence Architecture with Local LLMs for Building Automation} 


\author[1]{\fnm{Zhenghao} \sur{Wu}}\email{Zhenghao.Wu@ucdconnect.ie}

\author[1]{\fnm{Shaoshu} \sur{Zhu}}\email{Shaoshu.Zhu1@ucdconnect.ie}

\author[2]{\fnm{Shuyi} \sur{Sun}}\email{sun28@seas.upenn.edu}

\author[2]{\fnm{Zhiyuan} \sur{Chen}}\email{chennn@seas.upenn.edu}

\author[3]{\fnm{Dina} \sur{Shi}}\email{ds2489@cornell.edu}

\author[4]{\fnm{Chaofan} \sur{Li}}\email{Chaofan.Li@ucdconnect.ie}

\author*[1]{\fnm{Hadi Tabatabaee} \sur{Malazi}}\email{Hadi.TabatabaeeMalazi@ucd.ie}

\affil[1]{\small \orgdiv{School of Computer Science}, \orgname{University College Dublin}, \orgaddress{\city{Dublin}, \country{Ireland}}}

\affil[2]{\small \orgdiv{School of Engineering and Applied Science}, \orgname{University of Pennsylvania}, \orgaddress{\city{Pennsylvania}, \country{USA}}}

\affil[3]{\small \orgdiv{School of Electrical and Computer Engineering}, \orgname{Cornell University}, \orgaddress{\city{New York}, \country{USA}}}

\affil[4]{\small \orgdiv{Beijing Dublin International College}, \orgname{University College Dublin}, \orgaddress{\city{Beijing}, \country{China}}}


\abstract{
Building automation increasingly requires edge systems that can interpret ambiguous, multilingual, and context-dependent requests while checking structured outputs before device dispatch. We present SmartIntent, a serverless edge-intelligence architecture combining local large language model (LLM) intent services, device-state aggregation, a context-grounded JSON interface, deterministic pre-dispatch checks, orchestration, and device microservices. We adapt two compact LLMs with LoRA and evaluate them against compact, medium, and large local models and remote API baselines on a 2,000-instance Chinese-English-French test set. We distinguish offline structured-intent metrics from deployment-level validity, latency, and energy. The strongest adapted model, Qwen3.5-4B+LoRA, improves over its base model by 22.8, 20.9, and 13.1 percentage points in clarification handling, device identification, and action-parameter correctness, respectively. It matches or exceeds Qwen3.6-27B in format compliance and clarification handling and remains 1.0 and 1.9 percentage points below it in device and action-parameter correctness, respectively. In a local K3s/Knative deployment, it improves valid-output rate over its base model by 10.0--14.9 percentage points and reduces baseline-adjusted run-level energy by 5.2--18.4\%, with a latency increase of 10.9--16.6\%. For Qwen3.5-4B+LoRA, thinking mode adds at most 1.6 percentage points to output validity but increases latency by 58.2--118.2\% and energy by 78.5--152.2\%. These results show that compact local LLM services with deterministic output checks can support structured intent generation for execution-oriented building automation under controlled edge-deployment conditions while exposing clear validity-latency-energy trade-offs.
} 


\maketitle

\section{Introduction}
Modern buildings are increasingly equipped with networked sensors, controllable devices, and software services that can adapt indoor environments to occupant needs. As these environments become more connected, the role of automation is shifting from executing predefined device rules to supporting more flexible, intent-driven interaction in building automation~\cite{11331438}. In practice, occupants often express their needs through informal, incomplete, or context-dependent requests rather than precise automation commands. This creates a need for building automation systems that can interpret user intent in relation to available devices, current device-state context, and feasible control actions.

Supporting this form of interaction is technically challenging because user intent often cannot be interpreted from the utterance alone. Natural-language requests may be ambiguous, multilingual, incomplete, or expressed in terms of subjective goals, such as making a space more comfortable, brighter, or quieter. The same request may require different actions depending on the room, available devices, current device states, and device-control capabilities. Moreover, building automation outputs are not passive text responses: they must be translated into structured device actions and, where applicable, automation rules that are compatible with available devices and valid for downstream dispatch or rule management. The challenge, therefore, is to connect language understanding, context grounding, and executable automation in a form that remains suitable for edge-side operation.

Prior work has advanced several foundations relevant to LLM-enabled building automation, including serverless AIoT architectures, local LLM serving, intent-driven smart-environment interaction, and structured LLM output generation. Serverless and microservice-based AIoT architectures provide modular, event-driven, and resource-aware execution across cloud--edge--IoT environments~\cite{10767204}. Local and edge-hosted LLM serving studies investigate how to move language model inference closer to users, devices, and data sources under edge resource constraints~\cite{birkmose2025device}. In parallel, LLM-enabled smart-environment studies address natural-language automation and rule generation~\cite{10818469}, while structured-output methods improve schema-compliant LLM output generation~\cite {lu2025learning}. However, these research directions typically address individual aspects of the problem rather than the complete transition from user intent to executable AIoT actions. 
Limited attention has been given to grounding LLM-based interpretation in device context, converting it into structured automation representations, checking device/action/parameter constraints, and evaluating the resulting intent-to-execution workflow under edge-deployment constraints.

SmartIntent addresses this challenge through an intent-to-execution architecture
for LLM-enabled building automation, where language interpretation is treated as a deployable edge service rather than an external conversational capability. It coordinates local LLM intent services with context aggregation, structured intent representation, dispatching, and device-control microservices, while supporting rule management as part of the broader automation workflow. Rather than treating the LLM as a standalone conversational interface, SmartIntent embeds it as one service component within a structured automation workflow: user requests are interpreted against available device-state context, represented as machine-readable outputs, and checked for structural validity, device/action compatibility, and parameter feasibility before downstream dispatch and, where applicable, rule management. This design supports multilingual and ambiguity-aware interaction while keeping inference and orchestration within the local automation environment.  

A preliminary version of SmartIntent appeared in our CloudCom conference paper~\cite{11331438}. It introduced the core three-layer architecture, a prototype, a multilingual dataset, and an initial LoRA-based model evaluation. This journal article retains that foundation and includes the following revisions and substantive extensions:
\begin{itemize}
  \item Section~2 and Table~1 expand the related-work analysis and comparative positioning; Section~3 and Figure~1 refine the presentation of the previously published architecture.

  \item Section~4 formalizes the context-grounded JSON contract and deterministic pre-dispatch checks of schema, device, action, and parameter validity, and documents the heterogeneous K3s/Knative deployment. Figure~2, Tables~2--3, and Listings~1--3 are new.

  \item Sections~5.1--5.5 substantially extend the preliminary model study through hierarchical workload construction, LoRA adaptation of two 4B-class models, a held-out 2,000-instance Chinese--English--French test set, an explicit separation between deterministic checks and offline LLM-assisted assessment, and an updated suite of local and remote baselines evaluated under a common SmartIntent protocol. Figure~3 presents a substantially redesigned workload-generation pipeline; Figures~4--7 and Tables~4--8 present new configurations, experiments, and analyses that replace the preliminary conference evaluation. Listing~4 is also new.

  \item Sections~5.6--5.8 add a local deployment study of valid-output rate, response time, throughput, and baseline-adjusted run-level energy under default and context-stress workloads, with and without thinking mode. Figures~8--10 are new. 
\end{itemize}

The main contributions are as follows:
\begin{enumerate}
    \item A serverless edge-intelligence architecture for local LLM-enabled building automation, decomposing intent interpretation, context aggregation, orchestration, dispatching, and device-control functions into cooperating microservices.

    \item A context-grounded SmartIntent interface that represents natural-language instructions, language metadata, device-state context, device capabilities, clarification decisions, and machine-readable candidate device actions in a common JSON contract for downstream orchestration.
    
    \item An execution-oriented procedure for checking and evaluating LLM-generated automation outputs: deterministic pre-dispatch checks verify schema, device, action, and parameter validity, while targeted offline LLM-assisted assessment evaluates language consistency and instruction understanding.

    \item A compact-LLM adaptation study for multilingual and ambiguity-aware intent interpretation using LoRA-adapted local models and execution-oriented evaluation.

    \item A deployment-level evaluation of local LLM intent services in a heterogeneous K3s/Knative edge cluster, measuring valid-output rate, response time, throughput, and energy consumption under default and context-stress automation workloads.

    \item A reproducible multilingual SmartIntent benchmark workload with device-state context, clarification cases, and structured intent-to-execution labels~\footnote{\texttt{https://github.com/ucd-soc2/smartintent}}.
\end{enumerate}

The rest of this paper is organized as follows. Section~\ref{sec:Literature} reviews related work. Section~\ref{sec:Proposed} presents the SmartIntent architecture. Section~\ref{sec:Data} defines the data interface and deployment design. Section~\ref{sec:Performance} describes workload preparation, compact-LLM adaptation, and the model-level and deployment-level evaluations. Section~\ref{sec:Con} concludes the paper.

\section{Related work}
\label{sec:Literature}

This section reviews prior work across the four research directions that shape SmartIntent: serverless AIoT architectures, local LLM serving, intent-driven smart-environment automation, and validation of structured LLM outputs. 

\subsection{Serverless and microservice architectures for resource-aware AIoT}

In intent-driven AIoT automation, sensing, context aggregation, inference, validation, rule management, and actuation have different invocation patterns, resource demands, and latency requirements. Serverless and microservice architectures are therefore relevant not only as deployment technologies but also as mechanisms to separate, scale, and independently monitor these stages. Serverless computing supports event-driven execution, on-demand resource allocation, and scale-to-zero behavior, while microservice architectures support modular decomposition, loose coupling, independent deployment, per-service scaling, and fault isolation. These properties are particularly valuable for AIoT systems that integrate intermittent sensing, context-dependent decision-making, and device actuation across the computing continuum. 

Prior work has explored several architectural directions in serverless AIoT. Serverless sensing and fog/edge processing localize IoT data handling and reduce the need for persistently allocated services~\cite{8822951}. Cloud--edge--IoT intelligence, collaborative AIoT platforms, and serverless federated learning extend this model toward distributed training, inference, and application deployment across continuum resources~\cite{rong2021aiot,LOCONTE2024447,10767204}. Scalable AIoT testbeds and co-simulation frameworks further support evaluation of serverless execution under heterogeneous IoT workloads and deployment conditions~\cite{golec2025captain}. Together, these studies illustrate key architectural capabilities, including stateless execution, distributed service composition, cloud--edge--IoT orchestration, and scalable experimental support. Although these works do not address LLM-enabled building automation, they provide architectural patterns that inform SmartIntent. 

Energy-aware and autonomous scheduling in serverless edge environments provides another relevant foundation. Recent studies investigate how serverless functions can be scheduled across edge resources, accounting for residual energy, resource availability, and adaptive scheduler selection~\cite{ghorbian2025energy, tian2026autonomous,10.1145/3773274.3774929}. These works show that energy and resource sustainability are active concerns in serverless edge computing. However, they primarily focus on function placement, scheduler selection, or resource allocation, rather than decomposing LLM-based automation into cooperating microservices or evaluating energy/resource behavior across a multi-stage intent-to-execution pipeline.

Fewer studies examine how LLM-based intent interpretation can be transformed into an executable AIoT workflow, in which user requests are grounded in device context, represented as structured automation commands, checked against execution constraints, and evaluated as part of the automation system's operational behavior. SmartIntent addresses this gap by treating intent-driven building automation as a modular serverless AIoT pipeline in which each functional stage can be deployed, monitored, and evaluated as part of an executable command-to-actuation workflow. 

\subsection{Local LLM serving and runtime efficiency at the edge}
While serverless AIoT studies primarily address service decomposition and
resource orchestration, local LLM serving studies focus on the computational
characteristics and deployment efficiency of the inference workload itself. Local and edge-hosted LLM inference is increasingly relevant for AIoT automation because building-control prompts may contain device inventories, room-level states, occupancy patterns, behavioral routines, and comfort preferences~\cite{birkmose2025device}. Relying only on external cloud LLM services can also introduce network-dependent latency, availability risks, monetary cost, and limited control over deployment and energy/resource behavior. It may also raise privacy concerns when sensitive contextual information is processed outside the local automation environment~\cite{wang2026ssse}. For intent-driven building automation, these concerns are important because user commands must be interpreted quickly and converted into executable device actions or automation rules. Local LLM serving therefore provides a foundation for reducing cloud dependence while keeping inference close to the automation environment.

Recent systems work treats LLM inference as a runtime and serving problem. ServerlessLLM, for example, addresses low-latency serverless inference by optimizing model loading, checkpoint management, scheduling, and live migration for LLM workloads~\cite{fu2024serverlessllm}. Such work shows that LLM serving introduces challenges distinct from those of conventional stateless function execution, including checkpoint-loading delay, memory pressure, accelerator allocation, model scheduling, and startup latency. SmartIntent does not compete with such low-level LLM-serving runtimes. Instead, it examines how locally served LLM inference can be integrated as a measurable microservice within a broader AIoT automation pipeline. 

A related direction studies lightweight and on-device LLM inference for smart-home and edge environments. On-device home-assistant work shows that quantized LLMs can be deployed on constrained hardware for natural-language interaction and action generation, while exposing trade-offs between model precision, inference latency, classification accuracy, and hardware capability~\cite{birkmose2025device}. More broadly, efficient edge-LLM research examines techniques such as model compression, quantization, memory-efficient deployment, and resource-aware inference to reduce the cost of running LLMs under edge constraints~\cite{cai2026efficient}. Recent cloud--edge intelligence work also shows that partitioning tasks between lightweight edge components and larger reasoning models can reduce latency and data transmission while preserving task performance~\cite{chen2025synergistic}. These studies motivate the use of compact or locally hosted models in SmartIntent, where the LLM is not a monolithic assistant but a microservice whose latency and resource behavior affect downstream validation, dispatching, and rule execution. 

Energy and runtime efficiency are also central to local LLM deployment. Prior work on energy-aware LLM serving shows that inference cost depends strongly on workload characteristics, including input length, output length, model size, and hardware configuration~\cite{wilkins2024offline}. This is particularly relevant for intent-driven automation, where prompts may combine user commands, device metadata, current states, and contextual rules. Local LLM inference should therefore be evaluated not only as a functional component, but also as a workload whose per-request latency, throughput, resource footprint, and energy/resource cost depend on prompt and output characteristics. 

However, existing work on efficient LLM serving, on-device smart-home inference, and energy/runtime modeling usually treats inference as a standalone serving target, deployment optimization problem, or conversational assistant. Fewer studies jointly evaluate how local or edge-hosted LLM inference affects a context-aware, execution-oriented validation command-to-actuation pipeline under realistic automation workloads. SmartIntent addresses this gap by integrating local LLM inference as a measurable microservice within a serverless intent-to-execution architecture, enabling evaluation of its latency, throughput, resource footprint, fallback behavior, and energy/resource effects alongside the rest of the automation workflow. 

\subsection{LLM-enabled intent-to-execution in smart environments}

Smart environment automation is moving from explicit device-level commands toward intent-driven interaction, where users express desired outcomes rather than manually specifying each action. In building and smart-home environments,
such intents may be abstract, ambiguous, or context-dependent, and may also include multilingual expressions in practical deployments. They therefore require grounding in available devices, current states, environmental context, and, where available, user preferences and executable actions. Prior work has examined complementary aspects of this problem, including abstract-command interpretation~\cite{king2023getreadypartyexploring}, ambiguity in smart-environment commands~\cite{Calo_DeRussis_2024}, and disambiguation-oriented interaction~\cite{akasaki-sassano-2024-detecting}. These studies show that intent interpretation is not only a language-understanding task, but also a semantic-grounding problem that connects user expressions to device context and executable automation behavior.

A related body of work studies end-user automation and rule-based smart-home programming. Trigger-action platforms, such as IFTTT-like automation, allow users to define rules that connect events, conditions, and actions. However, prior studies show that users can face rule-creation and maintenance burdens, while rule ecosystems may suffer from conflicting applets, unsafe or unexpected behaviors, and difficulty managing multiple interacting automation rules~\cite{Yu_2021,255660,Huang_2019}. These limitations become more visible when users want to express abstract or context-rich goals rather than manually encode explicit trigger-action rules.

LLM-enabled smart-environment automation extends this direction in several functional ways. One line of work uses LLMs to generate, repair, or verify IoT automation rules from natural-language descriptions~\cite{10818469,Giudici2024}. Another line explores conversational or personalized smart-home automation, in which LLMs help users create routines or adapt automation behavior to user preferences and context~\cite{10975953,REYJOUANCHICOT2025101588}. A further line of studies focuses on on-device smart-home assistants and autonomous LLM agents for AIoT control~\cite{birkmose2025device,10729865}. Together, these studies shift smart-environment automation from manually specified rules toward language-mediated rule construction, adaptive routines, and agentic or on-device action generation. However, these directions differ in whether they produce executable rules, support immediate device actuation, rely on cloud-hosted LLM services, or evaluate integration with runtime automation infrastructure.

Many existing LLM-enabled systems therefore emphasize natural-language interaction, command disambiguation, rule generation, personalization, or autonomous control. Comparatively less attention has been given to how ambiguous or multilingual user input is grounded in live device context, represented structurally, and carried through to dispatchable device actions or persistent automation rules inside an orchestrated AIoT architecture. This distinction matters because a plausible LLM response is not enough for building automation: the output must be grounded in available devices, supported actions, current context, and the automation system's execution interfaces. 

SmartIntent addresses this gap by treating LLM-based interpretation as a context-aware intent-to-execution pipeline. It operationalizes this process through cooperating services that connect context aggregation and prompt construction to structured intent generation, clarification handling, dispatching, and rule management. In this way, SmartIntent sits between natural-language smart-home interfaces and runtime AIoT orchestration: it uses LLMs for semantic flexibility but constrains their outputs through context grounding, structured representation, and execution-oriented service integration. While intent interpretation and automation generation enable flexible interaction, their outputs must also satisfy structural and operational constraints before they can be safely incorporated into automation workflows.

\subsection{Validation and evaluation of structured LLM outputs}

In intent-driven AIoT automation, LLM outputs are not only textual responses; they may trigger device actions, create persistent automation rules, or request user clarification. This makes validation more critical than in ordinary text-generation settings. Errors in JSON format, device grounding, action selection, parameter values, or clarification flags can affect downstream execution. Syntactic correctness alone is also insufficient: a command may be schema-valid but semantically wrong, plausible but unsupported, or non-executable in the current automation context. 

For executable automation, structured-output generation and schema compliance are necessary because downstream services require machine-readable commands rather than free-form text. This literature addresses the difficulty of producing outputs that satisfy predefined schemas while preserving the intended content. Schema-based generation and reinforcement-learning approaches show that even modern LLMs can struggle with complex structured-output constraints~\cite{lu2025learning}. Post-processing approaches such as SLOT further demonstrate that unstructured LLM responses can be transformed into schema-compliant outputs in a model-agnostic manner~\cite{wang2025slot}. These studies motivate the need for schema-valid and machine-readable outputs, but they should be viewed as enabling techniques for runtime validation rather than direct runtime-safety mechanisms for AIoT execution.

A complementary direction concerns semantic assessment, including LLM-as-a-judge methods. Such methods offer scalable semantic assessment, but raise concerns about reliability, consistency, bias, and standardization~\cite{gu2026survey}. Conversely, exact matching and rule-based checks are more reproducible, but exact matching can be too rigid for ambiguous, paraphrased, or semantically equivalent automation commands. For AIoT automation, a hybrid validation strategy is therefore more appropriate. Deterministic checks verify objective execution properties, while LLM-assisted assessment is reserved for bounded semantic alignment between the user request and the generated command rather than determining execution feasibility. 

Existing work improves structured generation, schema compliance, and scalable semantic evaluation, but validation is usually studied as an offline evaluation, generation, or post-processing problem. Fewer studies embed hybrid validation as a constraint check for executable AIoT commands, especially when outputs must be checked before reaching device dispatch or rule execution. Such checks may include structural validity, semantic alignment, device/action compatibility, parameter feasibility, and clarification consistency. This runtime-gating perspective is important because an invalid LLM output in building automation is not merely an incorrect answer. It can become an incorrect, violating system action.

SmartIntent addresses this gap through execution-oriented output validation,
where generated commands are checked using deterministic constraints and
LLM-assisted semantic assessment before reaching downstream execution services. Deterministic checks verify schema conformance, required fields, device/action feasibility, parameter ranges, and clarification consistency. LLM-assisted assessment is used only when semantic judgment is required, such as checking whether a structured command aligns with the user's intent. In this way, SmartIntent treats validation as a runtime control mechanism rather than merely an offline evaluation step, so that outputs that fail the required structural, grounding, or feasibility checks are blocked, corrected, or routed for clarification before they affect the automation workflow. 

\subsection{Positioning of SmartIntent}

The reviewed studies demonstrate significant progress in individual aspects of
LLM-enabled AIoT automation, including edge deployment, language-based
interaction, structured output generation, and runtime efficiency. However,
existing approaches typically address these aspects separately and provide
limited support for an end-to-end execution workflow in which user intent is
interpreted, grounded in device context, transformed into structured automation
representations, checked against device, action, and parameter constraints, and
evaluated within an AIoT runtime environment. SmartIntent addresses this gap
through an execution-oriented serverless microservice architecture for
intent-driven AIoT automation. It treats LLM-based command interpretation not
only as a language-understanding task but as a measurable command-to-actuation
workflow, combining local LLM inference, context-aware prompt construction,
device-state aggregation, structured-output checking, dispatcher/rule-engine
integration, and runtime monitoring under automation workloads.  

\begin{table*}[t]
\footnotesize
\centering
\caption{Positioning of SmartIntent against representative related work.
S/M-AIoT: serverless or microservice AIoT architecture; L-LLM: local or
edge-hosted LLM inference; I2E: context-aware intent-to-execution; OVC:
Output Validation Before Control; RPE: runtime performance evaluation; ERE: energy/resource
evaluation.}
\label{tab:smartintent_positioning}
\renewcommand{\arraystretch}{1.10}
\setlength{\tabcolsep}{3pt}
\begin{tabularx}{\textwidth}{p{0.52\textwidth}cccccc}
\toprule
\textbf{Work / Direction} &
\textbf{S/M-AIoT} &
\textbf{L-LLM} &
\textbf{I2E} &
\textbf{OVC} &
\textbf{RPE} &
\textbf{ERE} \\
\midrule

IFTTT / end-user automation~\cite{255660,Yu_2021,Huang_2019}
& $\times$ & $\times$ & $\circ$ & $\circ$ & $\times$ & $\times$ \\

AutoIoT~\cite{10818469}
& $\times$ & $\times$ & $\checkmark$ & $\circ$ & $\circ$ & $\times$ \\

Green IFTTT~\cite{Giudici2024}
& $\times$ & $\times$ & $\circ$ & $\times$ & $\times$ & $\times$ \\

Lightweight/adaptive LLM smart-home systems~\cite{10975953}
& $\times$ & $\circ$ & $\checkmark$ & $\circ$ & $\circ$ & $\times$ \\

On-device home-assistant LLMs~\cite{birkmose2025device}
& $\times$ & $\checkmark$ & $\checkmark$ & $\circ$ & $\circ$ & $\times$ \\

Autonomous LLM agents for AIoT~\cite{10729865}
& $\times$ & $\times$ & $\checkmark$ & $\circ$ & $\circ$ & $\times$ \\

Serverless AIoT architecture/testbed studies~\cite{8822951,LOCONTE2024447,10767204,golec2025captain}
& $\checkmark$ & $\times$ & $\times$ & $\times$ & $\circ$ & $\circ$ \\

Energy-aware serverless edge scheduling~\cite{ghorbian2025energy,tian2026autonomous}
& $\checkmark$ & $\times$ & $\times$ & $\times$ & $\circ$ & $\checkmark$ \\

Serverless LLM serving~\cite{fu2024serverlessllm}
& $\circ$ & $\checkmark$ & $\times$ & $\times$ & $\checkmark$ & $\circ$ \\

Structured-output methods~\cite{lu2025learning,wang2025slot}
& $\times$ & $\times$ & $\times$ & $\circ$ & $\times$ & $\times$ \\

LLM-as-a-judge evaluation~\cite{gu2026survey}
& $\times$ & $\times$ & $\times$ & $\circ$ & $\times$ & $\times$ \\

\textbf{SmartIntent}
& $\checkmark$ & $\checkmark$ & $\checkmark$ & $\checkmark$ & $\checkmark$ & $\checkmark$ \\

\bottomrule
\multicolumn{7}{p{\textwidth}}{\footnotesize
The table summarizes each work or direction: $\checkmark$ = explicitly implemented or evaluated; $\circ$ = partially supported, indirectly addressed, or used as an enabling technique; $\times$ = not a main focus.}
\end{tabularx}
\end{table*}

In Table~\ref{tab:smartintent_positioning}, context-aware intent-to-execution refers to mapping natural-language user input, together with device or environmental context, into executable device actions or automation rules. Output validation refers to checking generated outputs before dispatching to downstream devices or executing rules. Runtime performance evaluation covers measured latency, throughput, or execution overhead; energy/resource evaluation covers measured or modeled energy and resource behavior. 

\section{Proposed approach}
\label{sec:Proposed}
We introduce \textit{SmartIntent}, a serverless microservice architecture designed for natural-language-driven control and orchestration of AIoT-enabled environments. Built on Knative, \textit{SmartIntent} combines lightweight LLM capabilities with modular microservices to support responsive interaction, ambiguity resolution, and context-aware device orchestration. 

SmartIntent adopts a three-layer edge-intelligence architecture that separates intent interpretation, context management, and device execution. Unlike conventional automation platforms where natural-language interpretation and device control are tightly coupled, SmartIntent decomposes the intent-to-execution workflow into independently deployable services. This layered design supports scalable deployment, service isolation, and runtime evaluation of performance and energy behavior under edge-cloud deployment constraints.

\begin{figure}[htbp!]
  \centering
  \includegraphics[width=0.8\textwidth]{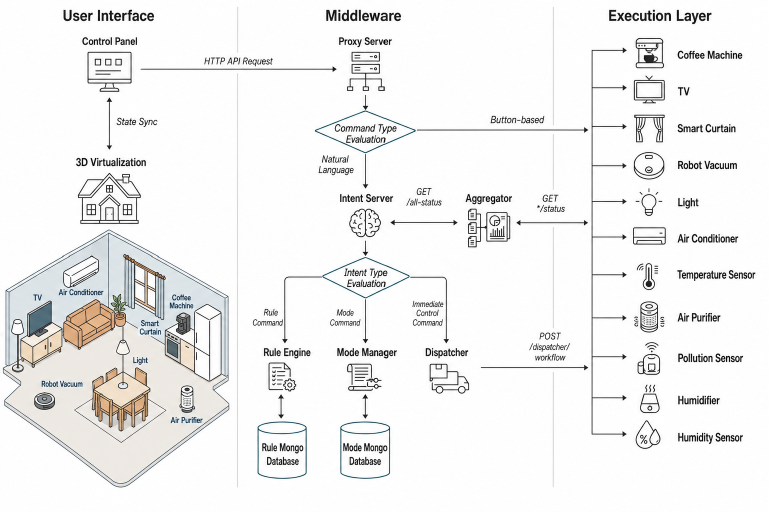}
  \caption{Overall architecture of SmartIntent.}
  \label{fig:architecture}
\end{figure}

\subsection{User interface layer}

The User Interface Layer provides two complementary mechanisms for interacting with the environment. First, a web-based control panel enables intuitive device management, with multilingual support via dynamic language switching. Performance optimizations such as input debouncing and conditional rendering reduce overhead while preserving responsiveness. Second, a 3D visualization module constructs spatial representations of rooms and devices. The visualization component shows device states spatially and supports interaction with the automation environment.

\subsection{Middleware layer}

The Middleware Layer is the cognitive core of SmartIntent, translating high-level user intents into executable operations. It comprises multiple stateless microservices deployed on a Knative-based serverless platform and managed through Kubernetes. All services expose RESTful endpoints for state reporting and control-portal integration. 

\textbf{1) Proxy Server} acts as the ingress gateway, routing all incoming interactions to the appropriate backend services. It provides a unified access layer for multiple input modalities (natural language commands and structured UI controls). The server forwards requests to the Intent Server, Aggregator, or device microservices. This design simplifies orchestration and enables loose coupling between the UI and middleware.

\textbf{2) Intent Server} performs semantic parsing of user commands. It first queries the Aggregator for device context, then constructs a prompt and invokes the LLM service to resolve the intent. The LLM service is deployed locally to enable edge-side inference and evaluate runtime behavior, and the output is returned either to the Dispatcher for immediate execution or to the Rule Engine for persistent automation. This process supports feedback and exception handling throughout. Let $u$ be the user utterance with language tag $\ell$,
and $s_t \in \mathcal{S}$ be the current device-state snapshot.
The intent parser parameterized by $\theta$ maps
\begin{equation}
f_\theta: (u, s_t) \longrightarrow
\hat{y} = (\hat d, \hat a, \hat{\mathbf p}, \hat m, \hat c),
\end{equation}
 where $\hat d \in \mathcal{D}$ is the device,
$\hat a \in \mathcal{A}(\hat d)$ is the action,
$\hat{\mathbf p}$ are typed parameters,
$\hat m$ is a natural-language summary,
and $\hat c \in \{0,1\}$ flags \texttt{needsClarification}.

\textbf{3) Aggregator} fuses real-time state data from all device microservices into a unified snapshot. It gathers the updates in parallel, with each request wrapped in a fault-tolerant handler to isolate failures. Responses are normalized into a structured object that represents the current system context, thereby supporting intent parsing, rule evaluation, and real-time visualization. 

\textbf{4) Dispatcher} coordinates command execution by routing structured control messages to device microservices. It decouples high-level intent processing from low-level device actuation, validates identifiers, enforces parameter integrity, and manages sequencing for multi-step workflows. Both atomic commands and compound chains are supported through a task queue with deterministic ordering. Fault isolation ensures that single-device failures do not disrupt broader workflows. 

\textbf{5) Rule Engine} applies event-condition-action (ECA) logic to enable autonomous behavior from contextual triggers. Rules pair Boolean conditions with actuation commands and can be managed. The engine periodically polls the Aggregator for state data, evaluates rules in batches with short-circuit optimization, and dispatches actions through the Dispatcher. This closes the control loop by enabling reactive automation and reducing the need for explicit user intervention. Based on ECA semantics, each rule $r=(\varphi,\alpha)$ consists of a Boolean condition
$\varphi:\mathcal{S}\times\mathcal{E}\!\to\!\{0,1\}$
over state $s_t$ and event stream $e_t$,
and an action $\alpha$. At time $t$,
\begin{equation}
\texttt{if } \varphi(s_t,e_t)=1 \texttt{ then } \mathrm{dispatch}(\alpha)
\end{equation}
With a rule set $\mathcal{R}$ and priority $\pi$, the executed action set is
\begin{equation}
\mathcal{A}_t=\{\alpha_r:\ r\in\mathcal{R},\ \varphi_r(s_t,e_t)=1\ \wedge\ \pi(r)\ \text{passes}\}
\end{equation}
State evolves as $s_{t+1}=F(s_t,\mathcal{A}_t,\varepsilon_t)$ with feedback $\varepsilon_t$ from sensors.

\textbf{6) Mode Manager} orchestrates predefined scenes by bundling multiple device actions into a single mode (e.g., ``Home'' or ``Night''). Modes are configured and stored in a database, then executed as sequential workflows via the Dispatcher. Database connectivity is verified before accepting requests to ensure reliability. This abstraction simplifies routine tasks, supports personalized automation, and enables one-shot execution of complex multi-device behaviors. 

These middleware services establish a cohesive processing pipeline that enables intelligent, intent-aware, and resilient control of the smart building system.

\subsection{Execution layer}
The Execution Layer converts high-level intents into device operations and feedback. It contains simulated smart devices and sensors as stateless microservices.

\textbf{1) Device and Sensor Services} offer standardized APIs for issuing control commands and querying states. Each device, such as lights, TVs, or air conditioners, and each sensor, such as temperature or air quality, is encapsulated in a separate microservice. Devices update their internal states when commands are executed, while sensors periodically refresh their readings for rule evaluation and UI updates. This modular design ensures seamless integration with upstream middleware components. 

\textbf{2) State consistency with Redis Integration} maintains operational continuity and recoverability. Each device microservice stores its current state (e.g., brightness, power mode, temperature setpoint) in a centralized Redis key-value store. Upon initialization, services retrieve their latest state from Redis, enabling state restoration after service restart and preserving current device context. This shared state store supports real-time queries by upstream components such as the Aggregator and Rule Engine. Redis was selected for its low-latency access, atomic operations, and suitability for time-sensitive control in distributed smart building environments.

\textbf{3) Environment--Device Feedback Simulation} models deterministic environmental changes based on device activity. For instance, the temperature sensor reflects cooling effects influenced by AC usage, fan speed, and mode. These feedback loops eliminate randomness, supporting temporal reasoning and stable control testing. This enables closed-loop simulation of perception--actuation interactions. 

\section{Data interface and deployment design}
\label{sec:Data}
This section defines how SmartIntent connects LLM-based language interpretation to device execution. Section~\ref{subsec:data-interface} specifies the JSON contract for requests, household context, clarification decisions, and candidate device actions. Section~\ref{subsec:deployment-and-implementation} describes the common service interface, edge-cluster environment, and experimental controls used to compare LLM backends under a consistent orchestration path.

\subsection{Data interface for intent orchestration}
\label{subsec:data-interface}

SmartIntent uses a fixed JSON interface to connect natural-language intent parsing with downstream device orchestration. The interface specifies the request submitted by the Intent Server, the household context supplied to the LLM backend, and the structured response returned for validation. The runtime dispatches only outputs that satisfy this contract.

Two payload schemas define the contract. The input combines the user instruction, language identifier, and household state. The output reports whether clarification is required and contains a user-facing message and a list of candidate device actions. The \texttt{message} field supports user communication, whereas the validated \texttt{results} array provides the machine-consumable commands for downstream execution.

\subsubsection{Intent input payload and embedded household state}

The intent input contains three top-level fields: \texttt{instruction}, \texttt{current\_state}, and \texttt{language}. The \texttt{instruction} field stores the user's natural-language request. The \texttt{language} field identifies the request language for multilingual intent processing and evaluation. The \texttt{current\_state} field represents the runtime household context as a flat list of device objects.

Each device object records its identifier, name, type, room, current status, adjustable parameters, and capability schema. The \texttt{status} field is a scalar state indicator, such as \texttt{on}, \texttt{off}, \texttt{open}, \texttt{closed}, \texttt{active}, \texttt{folded}, or \texttt{null}. Device-specific numeric or categorical values are stored in \texttt{parameters}. For example, a light may report its brightness, whereas an air conditioner may report its temperature or operating mode.

The \texttt{capabilities} field defines the control space each device exposes. It lists supported actions, numeric parameter ranges, and categorical parameter values. For an executable intent to be valid, its device, action, and parameter values must satisfy these declared constraints.

\begin{lstlisting}[style=jsonstyle, caption={SmartIntent input payload.}, label={lst:intent-input-payload}]
{
  "instruction": "Set the bedroom light brightness to 70 percent.",
  "language": "en",
  "current_state": [
    {
      "deviceId": "BedroomLight",
      "name": "bedroom light",
      "type": "light",
      "room": "bedroom",
      "status": "on",
      "parameters": {
        "brightness": 40
      },
      "capabilities": {
        "actions": ["turn_on", "turn_off", "set_brightness"],
        "numeric_parameters": {
          "brightness": [0, 100]
        }
      }
    }
  ]
}
\end{lstlisting}

This embedded state grounds language interpretation in the current device inventory and its declared capabilities. It also gives the runtime an explicit basis for determining whether a requested operation is executable or requires clarification.

\subsubsection{Intent output payload}

The LLM output must follow the SmartIntent schema, which contains \texttt{needsClarification}, \texttt{message}, and \texttt{results}. The \texttt{needsClarification} field indicates whether the request can be resolved as an executable or informational intent. The \texttt{message} field provides the natural-language response shown to the user. The \texttt{results} array contains candidate device commands and is the only portion eligible for downstream execution.

\begin{lstlisting}[style=jsonstyle, caption={SmartIntent output payload.}, label={lst:intent-output-payload}]
{
  "needsClarification": false,
  "message": "Okay, I will set the bedroom light brightness to 70 percent.",
  "results": [
    {
      "deviceId": "BedroomLight",
      "action": "set_brightness",
      "status": "on",
      "parameters": {
        "brightness": 70
      }
    }
  ]
}
\end{lstlisting}

The execution layer therefore consumes only validated objects from the \texttt{results} array. Keeping \texttt{message} outside the dispatch path prevents user-facing language from being interpreted as a device command.

\subsection{Deployment setup and implementation details}
\label{subsec:deployment-and-implementation}

We deploy the LLM backends as intent servers in our local edge cluster to evaluate their performance and energy efficiency. This deployment provides direct measurements of model behavior, service latency, resource use, and power consumption under a common orchestration path.

\subsubsection{Deployment setup}
The deployment ran on a heterogeneous Kubernetes edge cluster managed by K3s. Knative Serving orchestrated both the LLM intent services and smart-home device services. The cluster combines ARM, Apple Silicon, and x86 worker nodes with different GPU capabilities, as summarized in Table~\ref{tab:cluster}.

\begin{table}[ht!]
    \footnotesize
    \centering
    \begin{tabular}{ccccc}
    \toprule
        \textbf{Node ID} & \textbf{Role} & \textbf{CPU} & \textbf{GPU}& \textbf{DRAM (GB)} \\
    \specialrule{\heavyrulewidth}{3pt}{1pt}
    \specialrule{\heavyrulewidth}{0pt}{3pt}
        1 & Worker & AMD 5900X & RTX 3070 & 32 \\
        2 & Worker & Intel i5-10400f & GTX 1660 &32  \\
        3 & Worker & ARM Cortex-A78AE & Embedded &8  \\
        4 & Worker & ARM Cortex-A78AE & Embedded & 8   \\
        5 & Worker & Apple M4 & Embedded &16 \\
        6 & Control Plane & ARM Cortex-A76 & / & 8 \\
    \hline
    \end{tabular}
    \caption{Hardware configuration of the heterogeneous cluster nodes.} 
    \label{tab:cluster} 
\end{table}

To collect real-time power consumption for these worker nodes, we use Shelly Plus Plug~\footnote{https://www.shelly.com/products/shelly-plus-plug-uk/} meters to capture and expose the power data to our monitoring stack, composed of Prometheus and Grafana. The monitoring interval for other service metrics was aligned with the power-sampling interval.

We profiled each worker’s idle power after installing the experimental platform.
Before every run, we set the processor governor on supported nodes to \texttt{performance}. Only one LLM intent service was evaluated at a time. Section \ref{subsec:deployment-evaluation-protocol} defines the idle-power correction and run-level energy calculation. 

\subsubsection{Service implementation}

The deployment design separates intent inference from device execution. Each LLM backend is exposed as an OpenAI-compatible intent service, while the smart-home devices are implemented as independent microservices. This separation lets the same device context and validation logic run across different intent services and hardware nodes.

All intent services were packaged as containers and deployed through Knative Serving with Kourier-based ingress routing. Each container exposed inference, health-check, and model-discovery endpoints on a common service interface. GPU-enabled deployments use CUDA-compatible NVIDIA containers where applicable, and model images are prepared for the corresponding node architecture before deployment.

The device layer comprised microservices for state inspection, control, and health monitoring. Each service maintains its current state and capability schema, including supported actions and valid parameter ranges. Before dispatch, the runtime validated each predicted intent against this schema to prevent unsupported actions and commands targeting non-existent devices.

The evaluation pipeline sent each request to one deployed intent service at a time and applied a shared \texttt{system prompt} schema across all models. This design held the deployment interface and orchestration path constant while varying the LLM backend.

To compare the compact open-source LLMs with our fine-tuned model, we have selected four quantized baselines to deploy in our cluster, as shown in Table~\ref{tab:models}. 
\begin{table}[!t]
    \footnotesize
    \centering
    \begin{tabular}{ccccc}
    \toprule
        \textbf{Model} & \textbf{Size} & \textbf{Quantization} & \textbf{Context Window} & \textbf{Reasoning}\\
    \specialrule{\heavyrulewidth}{3pt}{1pt}
    \specialrule{\heavyrulewidth}{0pt}{3pt}
        Gemma-4-E2B-it & 4B & Q4\_K\_M & 32,768 & on \& off \\
        Qwen3.5-4B & 4B & Q4\_K\_M & 32,768 & on \& off \\
        Phi-4-mini & 3.8B & Q4\_K\_M  & 16,384 & - \\
        Nemotron-4B & 4B & Q4\_K\_M & 4,096 & - \\ 
        \textbf{Qwen3.5-4B + LoRA} & 4B & Q4\_K\_M & 32,768 & on \& off \\
    \hline
    \end{tabular}
    \caption{Compact LLMs used in the local deployment evaluation. Configured context denotes the server allocation rather than the model’s native context window.} 
    \label{tab:models} 
\end{table} 

The proposed model was derived from \textit{Qwen3.5-4B} through task-specific fine-tuning and converted to the GGUF format using Q4\_K\_M quantization. It was served through an OpenAI-compatible chat-completion endpoint with a 32768-token context window, full GPU-layer offloading, and a single inference slot. For comparison, the pipeline also supported the original \textit{Qwen3.5-4B}, \textit{Gemma-4-E2B}, \textit{Phi-4-Mini}, and \textit{Nemotron-4B}. We evaluated thinking and non-thinking configurations separately where supported by the corresponding model. \vspace{0.1in}

\begin{lstlisting}[style=jsonstyle, caption={Representative Knative service manifest for an LLM intent service.}, label={lst:LLM-manifest}]
apiVersion: serving.knative.dev/v1
kind: Service

spec:
  template:
    metadata:
      annotations:
        ...
    spec:
      runtimeClassName: nvidia
      containerConcurrency: 1
      timeoutSeconds: 120
      ...
      containers:
        - image: <MODEL IMAGE>
          imagePullPolicy: IfNotPresent
          args:
            - --host
            - 0.0.0.0
            - --port
            - "8000"
            - --ctx-size
            - "16384"
            - --n-gpu-layers
            - "999"
            - --parallel
            - "1"
            - --metrics
            - --jinja
            - --chat-template-kwargs
            - '{"enable_thinking":<true/false>}'
      ...
\end{lstlisting}

\begin{figure}[ht!]

    \centering

    \includegraphics[width=0.95\linewidth]{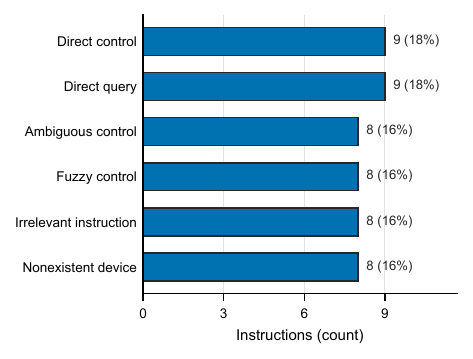}

    \caption{Instruction type distribution of \texttt{default} and \texttt{stress} workloads.}

    \label{fig:instruction-distribution}

\end{figure}

We evaluated the models using two Chinese-language workloads: \texttt{default} and \texttt{stress}, each containing 50 smart-home scenarios. The workloads covered direct control, direct query, fuzzy control, ambiguous control, nonexistent-device requests, and irrelevant instructions. We construct both workloads to have the same test case distribution, as illustrated in Fig.\ref{fig:instruction-distribution}. Each test case contained a natural-language instruction, the current states and capabilities of the available devices, and a reference output. The \texttt{default} workload represented scenarios with relatively short inputs and limited device context. The stress workload increased scenario complexity by providing longer inputs and larger contextual device-state descriptions, thereby increasing the number of input tokens the model processed. This design allowed us to examine whether the proposed fine-tuning remained effective as the amount of contextual information increased.

For each request, the model was instructed to return a JSON object containing a clarification indicator, a user-facing message, and a list of device actions. The system prompt restricted predictions to device identifiers, actions, and parameters present in the supplied context. Ambiguous instructions and nonexistent-device requests were labeled as clarification cases with empty executable results. The same prompt schema, workload order, and inference parameters were applied to all evaluated models.

An output was considered valid only if it could be parsed as JSON, conformed to the required schema, referenced existing devices, and contained only supported actions and parameters. We reported the valid-output rate as the proportion of requests satisfying all these requirements. Average response time was calculated from the elapsed time of individual requests. Aggregate completion throughput was calculated by dividing the total number of generated completion tokens by the wall-clock duration of the corresponding experimental run.

\section{Performance evaluation}
\label{sec:Performance}
This section evaluates SmartIntent at both the model and deployment levels. The model-level analysis tests whether task-specific fine-tuning improves compact LLMs on schema-sensitive and execution-oriented metrics. It also compares the adapted compact models with larger local and remote API baselines under a common SmartIntent input-output contract. The deployment-level analysis measures valid-output rate, response time, throughput, and energy consumption under different context loads and reasoning configurations.

\subsection{Experimental workload preparation and synthetic intent generation}
\label{subsec:data-preparation}

We evaluated SmartIntent in household-scale automation contexts. Each test case comprised a household state, a natural language instruction, and a reference intent output. The household state followed the data interface defined in Section~\ref{subsec:data-interface}. In the following, we describe how we generated the workload and prepared it for fine-tuning and evaluation. Table~\ref{tab:room-profile-control} summarizes the controlled sampling
dimensions used to construct the experimental household contexts.

The household-state payload described the parser's runtime environment, including rooms, devices, current states, and supported capabilities. The instruction expressed a control request, a state query, or a non-executable request. The expected output followed the SmartIntent schema and served as the reference for both supervised fine-tuning and evaluation.

\begin{figure}[t]
    \centering
    \includegraphics[width=0.95\linewidth]{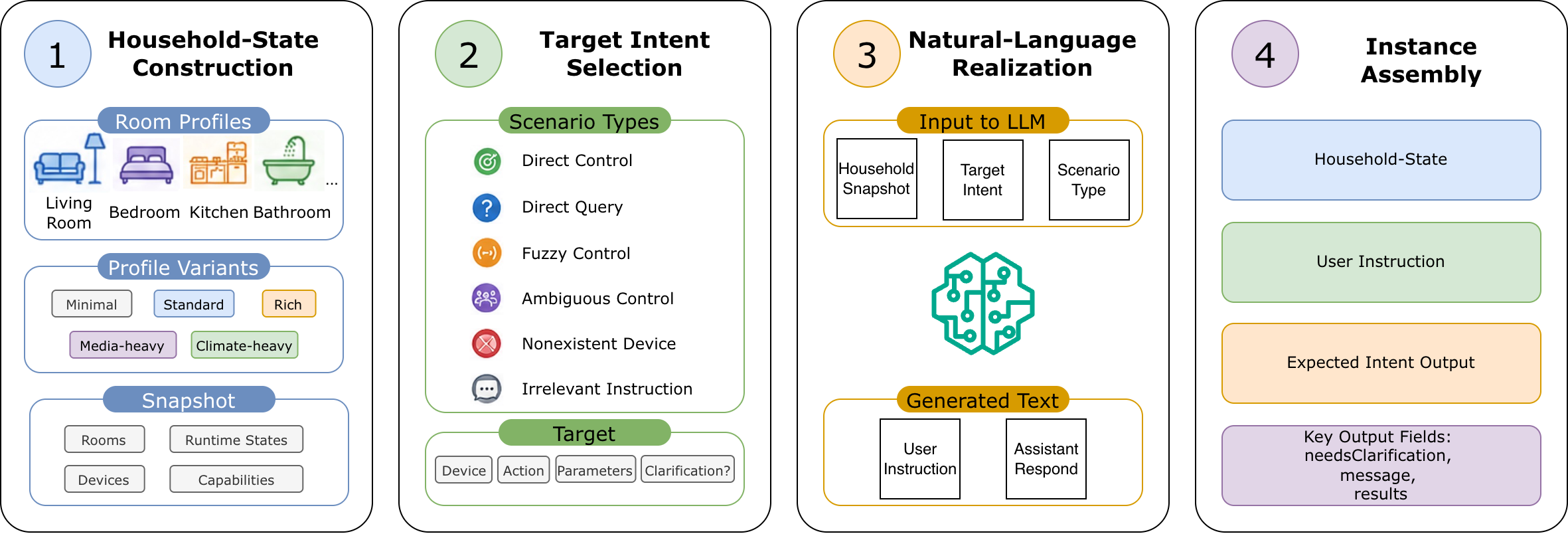}
    \caption{Hierarchical generation pipeline for SmartIntent data. The pipeline separates household state construction, target intent selection, natural language realization, and labeled instance assembly.}
    \label{fig:data-generation-pipeline}
\end{figure}

Figure~\ref{fig:data-generation-pipeline} presents the hierarchical synthesis pipeline. The pipeline constructed the household context and target intent before generating the corresponding natural language instruction. The household sampler defined the environment presented to the model. The target selector then determined the scenario, device, action, and intended state change. An LLM rendered this target as a natural language instruction, while the sampled state and target metadata determined the structured reference output. This separation introduced linguistic diversity without letting the LLM choose the ground-truth device, action, parameter, or clarification label.

\subsubsection{Household state construction}

The first stage constructed a household snapshot from room profiles. Each profile specified the room category, device families, device density, and contextual difficulty represented by a case. The implementation included living rooms, bedrooms, kitchens, bathrooms, and studies. Each room was instantiated as a minimal, standard, rich, media-heavy, or climate-heavy configuration.

The room profile sampler generated diverse household contexts that conformed to the SmartIntent data interface. It controlled the number of rooms, device density, and potential ambiguity in each household state. For example, a living room could contain a television, lamps, speakers, curtains, and climate control devices. A bathroom could contain a light, fan, heater, water heater, and dehumidifier. These profiles produced coherent room-level contexts with plausible patterns of device co-occurrence.

The sampler first selected the rooms in a household snapshot. It then sampled a profile variant for each room and expanded the profile into concrete devices. Each device received a runtime state and was represented according to the interface in Section~\ref{subsec:data-interface}. This representation included its scalar status, adjustable parameters, and capability schema.

\begin{table}[t]
    \footnotesize
    \centering
    \caption{Sampling dimensions used to construct experimental household contexts.}
    \label{tab:room-profile-control}
    \begin{tabular}{lll}
    \toprule
    \textbf{Sampling layer} & \textbf{Controlled factor} & \textbf{Experimental effect} \\
    \specialrule{\heavyrulewidth}{3pt}{1pt}
    \specialrule{\heavyrulewidth}{0pt}{3pt}
    Room selection & Household scope & Single room vs. multi-room context \\
    Room variant & Device density & Minimal vs. rich configuration \\
    Device family & Distractor composition & Lights, speakers, climate devices \\
    Runtime state & Current device values & On/off, temperature, brightness \\
    Capability schema & Valid action space & Supported actions and parameters \\
    \specialrule{\heavyrulewidth}{3pt}{1pt}
    \end{tabular}
\end{table}

The capability schema constrained deterministic label construction by specifying the available devices, supported actions, and valid parameter ranges. A single household state could therefore support several intent cases. Multiple lights in one room enabled an ambiguity case, whereas a request for an absent projector created a nonexistent device case. An air conditioner with temperature and mode controls supported either direct control or fuzzy comfort-related requests.

\subsubsection{Scenario-driven intent selection}
After sampling a household state, a rule-based selector generated the target intent. It first selected a scenario type and then sampled the target device, action, and state change. These choices were constrained by the current household state and device capability schema. The workload covered six scenario types:

\begin{itemize}
    \item \textbf{Direct control}: the user clearly requests a specific device state change or setting adjustment.
    \item \textbf{Direct query}: the user asks about the current state of a specific device.
    \item \textbf{Fuzzy control}: the user gives an imprecise request that is mapped to a capability-grounded action and parameter change determined by the target selector.    
    \item \textbf{Ambiguous control}: the user gives a request that may refer to multiple matching devices and requires clarification.
    \item \textbf{Nonexistent device}: the user refers to a device that is not available in the current household state.
    \item \textbf{Irrelevant instruction}: the user instruction is unrelated to smart home device control or state query.
\end{itemize}

These scenarios tested whether the parser selected the appropriate processing path for each request. Control cases required correct device identification and valid action generation. Ambiguous, nonexistent device, and irrelevant cases instead required the parser to withhold device commands and return an appropriate response.

\subsubsection{Natural language realization and labeled instance assembly}

The LLM rendered the selected target intent as a natural language instruction and, when required, a user-facing response message. The executable fields, clarification decision, device identifiers, actions, and parameters were constructed deterministically from the household state, device metadata, and selected target. The LLM controlled only the surface wording, style, specificity, and language of the generated text.

Each case combined the serialized household state payload and instruction with its reference output. The output followed the SmartIntent schema and contained \texttt{needsClarification}, \texttt{message}, and \texttt{results} fields. For control requests, \texttt{results} identified the target device and action. This field remained empty for ambiguous, nonexistent device, and irrelevant requests.

We used the same instance format for fine-tuning and evaluation. During fine-tuning, the model learned to map each household state payload and instruction to the expected structured output. During evaluation, we compared its prediction with the reference using the metrics defined in Section~\ref{subsec:evaluation-metrics}.

The resulting workload provided capability-grounded supervision across household configurations, scenario types, and languages. We used it to fine-tune the compact models, conduct offline evaluation, and prepare requests for the local edge-cluster deployment experiment.

\subsection{Compact LLM fine-tuning}
\label{subsec:compact-llm-finetuning}

We fine-tuned two compact instruction-tuned LLMs for SmartIntent parsing. The adaptation targeted reliable schema-valid JSON generation, accurate grounding in the supplied household state, and an edge-deployable model size. Grounding encompassed device and action selection, parameter generation, and clarification handling.

The two deployment targets were \texttt{unsloth/Qwen3.5-4B} and \texttt{unsloth/gemma-4-E2B-it}, both compact models at approximately the 4B scale. Both had been instruction-tuned before adaptation; we did not fine-tune base pretrained models. The experiment therefore examined limited supervision domain adaptation of compact instruction-following models.

The training set contained 1,000 Chinese SmartIntent instances generated using the process described in Section~\ref{subsec:data-preparation}. We formatted each instance as a three-message conversation. The system message defined the parser role and JSON schema constraint. The user message contained the instruction and serialized household state, and the assistant message provided the reference SmartIntent output.

\begin{lstlisting}[style=jsonstyle, caption={Fine-tuning instance format.}]
[
  {
    "role": "system",
    "content": "You are a smart home intent parser. Return only JSON matching the required schema."
  },
  {
    "role": "user",
    "content": "Return only a valid JSON object matching the required schema."
    Input JSON:{"instruction": "...", "current_state": [...], "language": "zh"}
  },
  {
    "role": "assistant",
    "content": "{"message": "...", "needsClarification": false, "results": [...]}"
  }
]
\end{lstlisting}

The assistant output followed the SmartIntent schema, with three top-level fields: \texttt{message}, \texttt{needsClarification}, and \texttt{results}. The \texttt{results} field was an array. In executable control cases, it contained one or more action objects. These objects included fields such as \texttt{deviceId}, \texttt{action}, \texttt{status}, and \texttt{parameters}. In clarification or non-executable cases, \texttt{needsClarification} was typically \texttt{true} and \texttt{results} was empty. Query cases could also set \texttt{needsClarification=false} with an empty \texttt{results} array when no device action was required.

We used the Unsloth framework for supervised fine-tuning. Both models were loaded in BF16/16-bit mode and adapted using low-rank adaptation (LoRA). This common setting matched the adaptation method and loading precision across models, avoiding a comparison between LoRA and quantized LoRA configurations. In both cases, we applied the tokenizer's chat template using \texttt{apply\_chat\_template}. We used no validation split during training and assessed the final models on a held-out test set.

\begin{table}[t]
    \footnotesize
    \centering
    \caption {Fine-tuning configurations for compact SmartIntent models.}
    \label{tab:finetuning-config}
    \begin{tabular}{lll}
    \toprule
    \textbf{Configuration} & \textbf{Qwen3.5-4B} & \textbf{Gemma-4-E2B-it} \\
    \specialrule{\heavyrulewidth}{3pt}{1pt}
    \specialrule{\heavyrulewidth}{0pt}{3pt}
    Model path & \texttt{unsloth/Qwen3.5-4B} & \texttt{unsloth/gemma-4-E2B-it} \\
    Loading mode & BF16 / 16-bit & BF16 / 16-bit \\
    Adaptation & LoRA & LoRA \\
    LoRA rank & 16 & 16 \\
    LoRA alpha & 16 & 16 \\
    LoRA dropout & 0.0 & 0.0 \\
    Max sequence length & 4096 & 8192 \\
    Batch size & 1 & 4 \\
    Gradient accumulation & 8 & 4 \\
    Epochs & 5 & 5 \\
    Learning rate & $1\times10^{-4}$ & $1\times10^{-4}$ \\
    Optimizer & \texttt{adamw\_8bit} & \texttt{adamw\_torch} \\
    Training hardware & NVIDIA L40S & NVIDIA L40S \\
    Final train loss & 0.123 & 0.175 \\
    \specialrule{\heavyrulewidth}{3pt}{1pt}
    \end{tabular}
\end{table}

After fine-tuning, the output directories contained LoRA adapter weights rather than merged full model checkpoints. During inference, the serving script loaded each base model with its trained LoRA adapter. This workflow avoided maintaining a separate merged checkpoint for each fine-tuned variant.

The held-out test set contained 2,000 instances: 668 Chinese, 666 English, and 666 French samples. Because fine-tuning used only Chinese instances, this multilingual set also tested whether schema adaptation preserved cross-lingual generalization. We evaluated the fine-tuned models offline and used them in the deployment workflow to measure runtime power. Table~\ref{tab:finetuning-config} summarizes the fine-tuning configurations used for the two compact SmartIntent models.
\subsection{Evaluation protocol and metrics}
\label{subsec:evaluation-metrics}

We evaluated SmartIntent as a structured intent orchestration component. Given input payload $x_i$, reference output $y_i$, and raw response $r_i$, the evaluator recovered and validated a structured prediction $\hat{y}_i$. If it could not obtain a schema-valid prediction, it marked $\hat{y}_i$ as invalid. The response then received a score of 0 for every applicable downstream metric.

For each test case $i$, the evaluator computed six metrics:
\begin{equation}
\mathbf{m}_i =
\left[
O_i, C_i, D_i, A_i, L_i, U_i
\right],
\label{eq:metric-vector}
\end{equation}
Here, $O_i$, $C_i$, and $D_i$ denote output format compliance, clarification handling, and device identification, respectively. The terms $A_i$, $L_i$, and $U_i$ denote action-parameter correctness, language consistency, and instruction understanding, respectively.

Let $c(y)$ denote the value of \texttt{needsClarification}, and let $R(y)$ denote the \texttt{results} list. The output mode is defined as
\begin{equation}
\operatorname{Mode}(y)=
\begin{cases}
\textsc{clarify},
& c(y)=\texttt{true} \land R(y)=\emptyset,\\
\textsc{query},
& c(y)=\texttt{false} \land R(y)=\emptyset,\\
\textsc{execute},
& c(y)=\texttt{false} \land R(y)\neq\emptyset,\\
\textsc{invalid},
& \text{otherwise}.
\end{cases}
\label{eq:output-mode}
\end{equation}
An output with \texttt{needsClarification=true} and a non-empty result list is therefore treated as invalid.

The evaluation set for each metric was
\begin{equation}
\mathcal{T}_O
=
\mathcal{T}_C
=
\mathcal{T}_D
=
\mathcal{T}_A
=
\mathcal{T}_L
=
\mathcal{T}_U
=
\mathcal{T}.
\label{eq:metric-test-sets}
\end{equation}
The reported score for metric $k$ was
\begin{equation}
M_k
=
\frac{1}{|\mathcal{T}_k|}
\sum_{i\in\mathcal{T}_k}
m_{i,k},
\label{eq:metric-score}
\end{equation}
where $m_{i,k}\in\{0,1\}$ is the value of metric $k$ for case $i$.

For each case, let $\mathcal{K}_i$ denote the set of applicable metrics:
\begin{equation}
K_i = \{O, C, D, A, L, U\}.
\label{eq:applicable-metrics}
\end{equation}
The per-case aggregate score was
\begin{equation}
S_i
=
\frac{1}{|\mathcal{K}_i|}
\sum_{k\in\mathcal{K}_i}
m_{i,k}.
\label{eq:aggregate-score}
\end{equation}

\paragraph{Output format compliance}
Output format compliance measured whether the raw response formed a standalone SmartIntent JSON payload. Let $\operatorname{parse}(r_i)$ denote the JSON recovery and schema validation procedure. We defined the metric as
\begin{equation}
O_i =
\mathbb{I}\left[
\operatorname{parse}(r_i)\neq\varnothing
\land
r_i\in\mathcal{J}_{\mathrm{standalone}}
\right],
\label{eq:format-compliance}
\end{equation}
where $\mathcal{J}_{\mathrm{standalone}}$ contained either a bare JSON object or one complete, independent fenced JSON block. We permitted no natural language text before or after the JSON payload. The parser could therefore recover a response with surrounding text even though that response failed the format compliance metric.

\paragraph{Clarification handling}
Clarification handling measured whether the model correctly distinguished clarification, non-executable query, and executable control outputs:
\begin{equation}
C_i =
\mathbb{I}\left[
\hat{y}_i\neq\varnothing
\land
\operatorname{Mode}(\hat{y}_i)
=
\operatorname{Mode}(y_i)
\right].
\label{eq:mode-correctness}
\end{equation}
This definition required query and clarification outputs to contain no executable results. It also prevented control commands from receiving credit in query cases.

\paragraph{Device identification}
Device identification measured whether the prediction followed the expected device-related output behavior for the reference output. For executable cases, it required the predicted and reference device sets to match exactly. For clarification cases, it required the prediction to request clarification and return no executable results. For query cases, it required the prediction not to request clarification. Thus,
\begin{equation}
D_i =
\begin{cases}
\mathbb{I}\!\left[
\hat{y}_i \neq \emptyset
\land
\mathrm{DevSet}(\hat{y}_i) = \mathrm{DevSet}(y_i)
\right],
& \mathrm{Mode}(y_i) = \mathrm{execute},
\\[4pt]
\mathbb{I}\!\left[
\hat{y}_i \neq \emptyset
\land
c(\hat{y}_i) = \mathrm{true}
\land
R(\hat{y}_i) = \emptyset
\right],
& \mathrm{Mode}(y_i) = \mathrm{clarify},
\\[4pt]
\mathbb{I}\!\left[
\hat{y}_i \neq \emptyset
\land
c(\hat{y}_i) = \mathrm{false}
\right],
& \mathrm{Mode}(y_i) = \mathrm{query}.
\end{cases}
\end{equation}

Here, $\mathrm{DevSet}(y)$ denotes the set of device identifiers appearing in $R(y)$.

\paragraph{Action-parameter correctness}
Action-parameter correctness measures whether the prediction followed the expected action behavior for the reference output. For executable cases, it checked the reference actions and parameters against the predicted results. For clarification cases, it required clarification with no executable result. For query cases, for which the reference contains no executable action, a successfully parsed prediction received credit under this metric.
\begin{equation}
\begin{aligned}
\operatorname{ResMatch}(q,e,x_i)
={}&
\left(q.\texttt{deviceId}=e.\texttt{deviceId}\right)\\
&{}\land
\left(q.\texttt{action}=e.\texttt{action}\right)\\
&{}\land
\operatorname{ParamMatch}(q,e,x_i).
\end{aligned}
\label{eq:result-match}
\end{equation}

The reference results were considered matched when each reference result had a corresponding predicted result:
\begin{equation}
\mathrm{ActMatch}(\hat{y}, y, x_i)
=
\forall e \in R(y),\;
\exists q \in R(\hat{y})
\text{ such that }
\mathrm{ResMatch}(q,e,x_i).
\label{eq:action-match}
\end{equation}
For each $i \in T$, action-parameter correctness was
\begin{equation}
A_i =
\begin{cases}
\mathbb{I}\!\left[
\hat{y}_i \neq \emptyset
\land
\mathrm{ActMatch}(\hat{y}_i, y_i, x_i)
\right],
& \mathrm{Mode}(y_i) = \mathrm{execute},
\\[4pt]
\mathbb{I}\!\left[
\hat{y}_i \neq \emptyset
\land
c(\hat{y}_i) = \mathrm{true}
\land
R(\hat{y}_i) = \emptyset
\right],
& \mathrm{Mode}(y_i) = \mathrm{clarify},
\\[4pt]
\mathbb{I}\!\left[
\hat{y}_i \neq \emptyset
\right],
& \mathrm{Mode}(y_i) = \mathrm{query}.
\end{cases}
\label{eq:action-parameter-correctness}
\end{equation}

The parameter matching function required identical predicted and reference parameter key sets. We compared non-numeric values by exact equality. Numeric values satisfied
\begin{equation}
|\hat{v}-v|
\leq
\tau_p,
\label{eq:numeric-tolerance}
\end{equation}
where $\tau_p$ was 0.5 for temperature, 5.0 for brightness, 5.0 for volume, and 3.0 for humidity. For all other numeric fields, $\tau_p=10^{-6}$. In fuzzy control cases, we also accepted values that changed in the reference direction relative to the current device state. We did not score the \texttt{status} field using this metric.

\paragraph{Language consistency}
Language consistency measured whether the user-facing response followed the language specified by the input payload. Let $\ell_i$ denote the target language. We defined the metric as
\begin{equation}
L_i =
\mathbb{I}\left[
\hat{y}_i\neq\varnothing
\right]
\cdot
\operatorname{Judge}_{\mathrm{lang}}
(x_i,y_i,\hat{y}_i,\ell_i),
\label{eq:language-consistency}
\end{equation}
where $L_i\in\{0,1\}$. This metric assessed the language of the \texttt{message} field independently of executable command correctness.

\paragraph{Instruction understanding}
Instruction understanding measured whether the prediction preserved the request's semantic intent under the supplied household state:
\begin{equation}
U_i =
C_i
\cdot
\operatorname{Judge}_{\mathrm{intent}}
(x_i,y_i,\hat{y}_i),
\label{eq:instruction-understanding}
\end{equation}
where $U_i\in\{0,1\}$. A prediction with an incorrect output mode received no credit, even if some content was semantically related to the request.

The evaluation combined deterministic interface checks with LLM-as-a-judge semantic checks. We computed output format compliance, clarification handling, device identification, and action-parameter correctness deterministically. The LLM judge assessed language consistency and instruction understanding. Section~\ref{subsubsec:llm-judge-config} specifies the judge model, reasoning mode, and common prompt. We computed device and action-parameter scores for all cases. For executable cases, they assessed device grounding and action-parameter correctness; for clarification cases, they also captured whether execution was correctly withheld. Query cases followed the special handling defined above.
\subsection{Model-level offline evaluation setup and baselines}
\label{subsec:model-eval-setup}

The offline evaluation compared general-purpose baselines with the fine-tuned compact models before deployment-level measurement. All models used the same held-out test set, input format, prompt template, and output schema.

\subsubsection{Offline inference setting}

Each model received the input payload defined in Section~\ref{subsec:data-interface} and was instructed to return only a SmartIntent JSON object.

We evaluated every model on the same held-out set of 2,000 instances: 668 Chinese, 666 English, and 666 French samples. We used the metrics defined in Section~\ref{subsec:evaluation-metrics}. Unless stated otherwise, each reported score covered the full test set. Language- and scenario-specific analyses assessed reliability across languages and intent types, particularly for fuzzy control and ambiguous requests.

We conducted offline inference on a Windows 11 workstation with an AMD Ryzen 9 9950X CPU, 64 GB memory, and an NVIDIA GeForce RTX 5090 GPU. Local models used a Llama. cpp-based backend exposed through an OpenAI-compatible interface. Remote API models used the same request format when the provider supported it. This shared invocation path placed local and remote backends under the same SmartIntent input-output contract.

\subsubsection{Prompting and reasoning mode control}

All baseline and fine-tuned models used the same SmartIntent prompt template. It specified the parser role, JSON-only constraint, and input/output fields defined in Section~\ref{subsec:data-interface}. We added no model-specific prompt engineering, except for reasoning mode controls required by some models or providers.

For models with explicit reasoning or thinking modes, we disabled these modes using the documented control for each model or backend. Depending on the backend, the control was an inference parameter or a prompt-level instruction. This setting tested each model as a structured intent parser that produced a directly parseable SmartIntent object, rather than explanatory or chain-of-thought-style text.

\subsubsection{Baseline model groups}

The baseline set covered several model scales and families. Because it did not form a complete family-by-size grid, we grouped models by approximate scale. We therefore did not interpret the comparison as a controlled scaling analysis. Instead, we tested whether task-adapted compact models approached larger general-purpose models under a common input-output interface.

We organized the models into four baseline groups. The compact local group comprised Gemma-4-E2B-it, Nemotron-4B, Phi-4-mini, Qwen3-4B, and Qwen3.5-4B. The medium local group comprised Gemma-4-E4B and Qwen3.5-9B. The large local group comprised Qwen3.6-27B and Gemma-4-31B. These groups spanned approximate scales of 4B, 8B--14B, and 27B--31B, respectively. The remote API group comprised DeepSeek-V4-Flash Preview and DeepSeek-V4-Pro Preview. We reported the fine-tuned models alongside these baselines to compare behavior before and after adaptation under the same evaluation contract.

\begin{table}[t]
\footnotesize
\centering
\caption{Model groups used in the offline model-level evaluation.}
\label{tab:baseline-models}
\begin{tabular}{llll}
\toprule
\textbf{Group} & \textbf{Model} & \textbf{Serving mode} & \textbf{Role in evaluation} \\
    \specialrule{\heavyrulewidth}{3pt}{1pt}
    \specialrule{\heavyrulewidth}{0pt}{3pt}
    Compact local & Gemma-4-E2B-it & Local & Pre fine-tuning baseline \\
Compact local & Nemotron-4B & Local & Compact baseline \\
Compact local & Phi-4-mini & Local & Compact baseline \\
Compact local & Qwen3-4B & Local & Compact baseline \\
Compact local & Qwen3.5-4B & Local & Qwen3.5 fine-tuning comparator \\
Medium local & Gemma-4-E4B & Local & Medium scale baseline \\
Medium local & Qwen3.5-9B & Local & Medium scale baseline \\
Large local & Qwen3.6-27B & Local & Large scale baseline \\
Large local & Gemma-4-31B & Local & Large scale baseline \\
Remote API & DeepSeek-V4-Flash Preview & API & Remote baseline \\
Remote API & DeepSeek-V4-Pro Preview & API & Remote baseline \\
Fine-tuned compact & Gemma-4-E2B-it + LoRA & Local & Fine-tuned SmartIntent model \\
Fine-tuned compact & Qwen3.5-4B + LoRA & Local & Fine-tuned SmartIntent model \\
\hline
\end{tabular}
\end{table}

The evaluation quantified the effect of supervised fine-tuning on each compact model and compared the adapted models with larger local and remote baselines.

\subsubsection{LLM-as-a-judge configuration}
\label{subsubsec:llm-judge-config}

We used the official April 2026 DeepSeek-V4-Flash Preview API model (\texttt{deepseek-v4-flash}) for both semantic metrics. We disabled thinking mode and requested a direct structured assessment. We evaluated every candidate model, language, scenario, and test instance with the same judge prompt. The prompt received the user input, reference output, and raw candidate response, and returned one Boolean decision per metric. Only language consistency and instruction understanding used these decisions; the other four metrics remained deterministic.

\subsection{Model-level evaluation results}
\label{subsec:model-eval-results}

We report offline results under the common protocol in Section~\ref{subsec:evaluation-metrics}.

\subsubsection{Overall comparison across model scales}

Table~\ref{tab:model-overall-results} compares the models across six interface-level metrics. The proposed Qwen3.5-4B + LoRA model outperformed its direct Qwen3.5-4B comparator on five metrics. It also approached the strongest large local baseline.

\begin{table*}[t]
\footnotesize
\centering
\caption{Overall model-level results on the held-out SmartIntent test set. Format, clarification, device, and action-parameter scores were computed deterministically. Language consistency and instruction understanding were assessed by the LLM judge.}
\label{tab:model-overall-results}
\begin{tabular}{lcccccc}
    \toprule
\textbf{Model} &
\textbf{Format} &
\textbf{Clarif.} &
\textbf{Device} &
\textbf{Act./Param.} &
\textbf{Lang.} &
\textbf{Intent} \\
    \specialrule{\heavyrulewidth}{3pt}{1pt}
    \specialrule{\heavyrulewidth}{0pt}{3pt}
    Gemma-4-E2B-it & 0.970 & 0.660 & 0.632 & 0.664 & 0.904 & 0.506 \\
Gemma-4-E4B-it & 0.950 & 0.835 & 0.814 & 0.825 & 0.878 & 0.790 \\
Gemma-4-31B & 0.999 & 0.891 & 0.873 & 0.864 & 0.917 & 0.799 \\
\hline
Qwen3-4B & 0.971 & 0.642 & 0.554 & 0.572 & 0.920 & 0.389 \\
Qwen3.5-4B & 0.988 & 0.640 & 0.635 & 0.720 & 0.986 & 0.589 \\
Qwen3.5-9B & 0.905 & 0.680 & 0.656 & 0.666 & 0.923 & 0.677 \\
Qwen3.6-27B & 0.996 & 0.862 & 0.854 & 0.870 & 0.981 & 0.832 \\
\hline
Nemotron-4B & 0.958 & 0.547 & 0.513 & 0.575 & 0.924 & 0.336 \\
Phi-4-mini & 0.965 & 0.479 & 0.450 & 0.611 & 0.385 & 0.338 \\
\hline
DeepSeek-V4-Flash Preview & 0.992 & 0.781 & 0.766 & 0.793 & 0.921 & 0.716 \\
DeepSeek-V4-Pro Preview & 0.994 & 0.767 & 0.747 & 0.774 & 0.900 & 0.698 \\
\hline
Qwen3.5-4B + LoRA (proposed) & 0.998 & 0.868 & 0.844 & 0.851 & 0.985 & 0.753 \\
\hline
\end{tabular}
\end{table*}

The proposed model achieved 0.998 format compliance, 0.868 clarification handling, 0.844 device identification, and 0.851 action-parameter correctness. The corresponding scores for Qwen3.6-27B were 0.996, 0.862, 0.854, and 0.870. The proposed model also achieved the highest language consistency score (0.985), but its instruction understanding score (0.753) remained below that of Qwen3.6-27B (0.832).

\begin{figure}[t]
    \centering
    \includegraphics[width=0.7\linewidth]{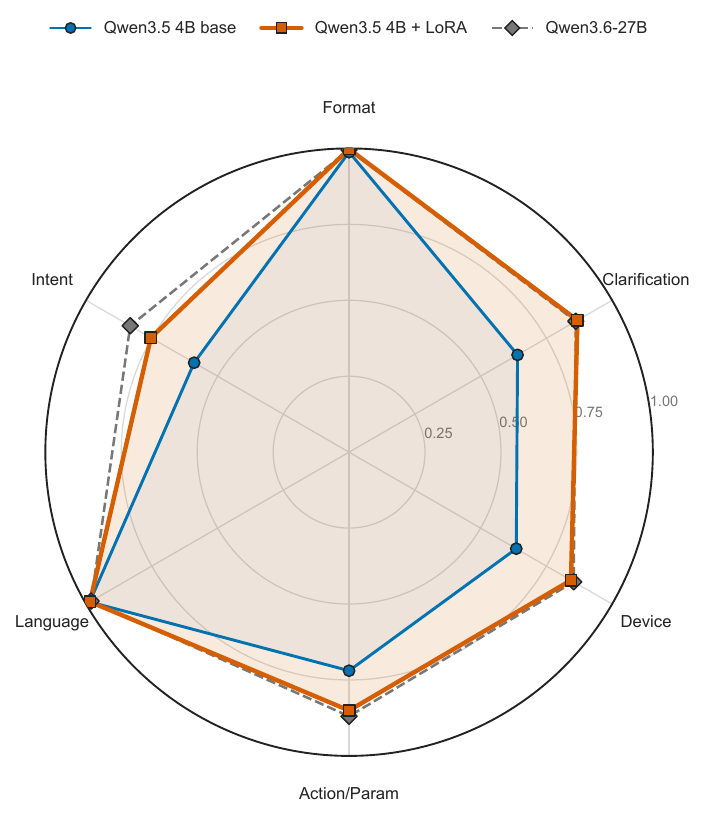}
    \caption{Model-level comparison between the Qwen3.5-4B base model, Qwen3.5-4B + LoRA, and Qwen3.6-27B across six SmartIntent metrics.}
    \label{fig:qwen-radar}
\end{figure}

Figure~\ref{fig:qwen-radar} compares the Qwen3.5-4B base model, its LoRA-adapted counterpart, and the larger Qwen3.6-27B baseline. This design shows the direct effect of Qwen3.5-4B fine-tuning while retaining a larger model reference. The fine-tuned compact model narrowed the gap to Qwen3.6-27B across the schema-sensitive metrics. It matched or exceeded the larger model on format compliance, clarification handling, and language consistency. The remaining deficits were 0.010 for device identification, 0.019 for action-parameter correctness, and 0.079 for instruction understanding.

\subsubsection{Effect of compact LLM fine-tuning}

To isolate the effect of fine-tuning, Table~\ref{tab:finetuning-effect} compares each adapted compact model with its original instruction-tuned checkpoint. This comparison tested whether supervised adaptation improved schema following, household state grounding, and action-parameter generation.

\begin{table}[t]
\footnotesize
\centering
\caption{Compact SmartIntent models before and after fine-tuning. The non-fine-tuned baselines used direct local inference. Format, clarification, device, and action-parameter scores were computed deterministically. Language consistency and instruction understanding were assessed by the LLM judge.}
\label{tab:finetuning-effect}
\begin{tabular}{lcccccc}
\toprule
\textbf{Model} &
\textbf{Format} &
\textbf{Clarif.} &
\textbf{Device} &
\textbf{Act./Param.} &
\textbf{Lang.} &
\textbf{Intent} \\
    \specialrule{\heavyrulewidth}{3pt}{1pt}
    \specialrule{\heavyrulewidth}{0pt}{3pt}
    Gemma-4-E2B-it & 0.970 & 0.660 & 0.632 & 0.664 & 0.904 & 0.506 \\
Gemma-4-E2B-it + LoRA & 0.916 & 0.569 & 0.554 & 0.685 & 0.837 & 0.402 \\
Qwen3.5-4B & 0.988 & 0.640 & 0.635 & 0.720 & 0.986 & 0.589 \\
Qwen3.5-4B + LoRA & 0.998 & 0.868 & 0.844 & 0.851 & 0.985 & 0.753 \\
\hline
\end{tabular}
\end{table}

\begin{figure*}[t]
    \centering
    \includegraphics[width=\textwidth]{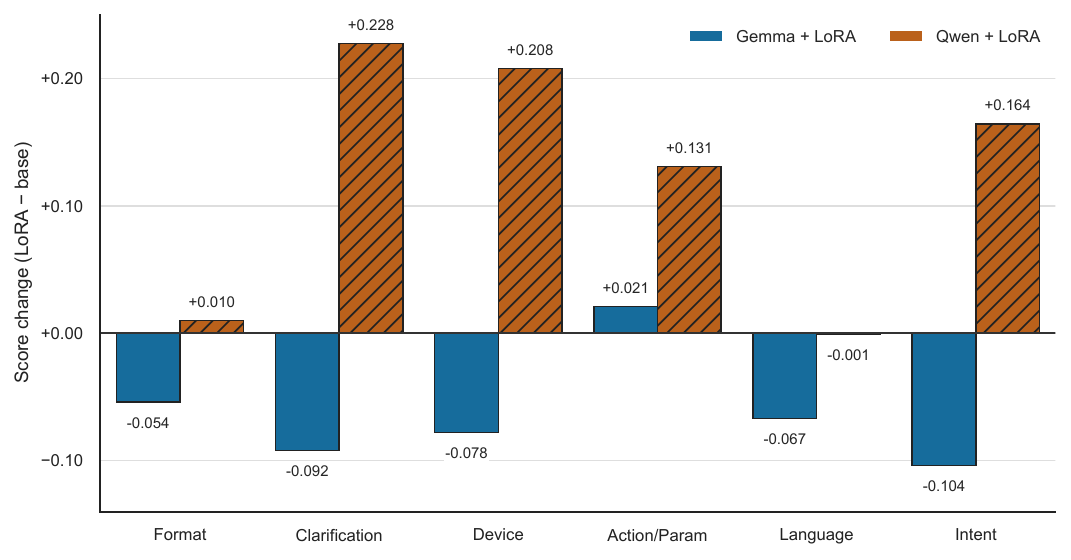}
    \caption{Effect of LoRA fine-tuning on compact SmartIntent models. Values denote fine-tuned score minus base model score.}
    \label{fig:finetuning-delta}
\end{figure*}

Figure~\ref{fig:finetuning-delta} summarizes the comparison as score deltas, calculated as the fine-tuned score minus the base model score. For Qwen3.5-4B, fine-tuning increased clarification handling by 0.228 and device identification by 0.209. Instruction understanding increased by 0.164, and action-parameter correctness increased by 0.131. Format compliance increased by 0.010, whereas language consistency decreased by 0.001.

Gemma-4-E2B-it showed the opposite pattern. Fine-tuning reduced format compliance by 0.054, clarification handling by 0.091, and device identification by 0.078. Language consistency and instruction understanding decreased by 0.067 and 0.104, respectively, whereas action-parameter correctness increased by 0.021. The format failures primarily involved analysis or explanatory text surrounding the JSON payload. The evaluator accepted bare JSON and complete standalone fenced blocks but rejected any additional text outside the structured payload.

\subsubsection{Multilingual and scenario-level analysis}

\begin{figure}[t]
    \centering
    \includegraphics[width=.95\linewidth]{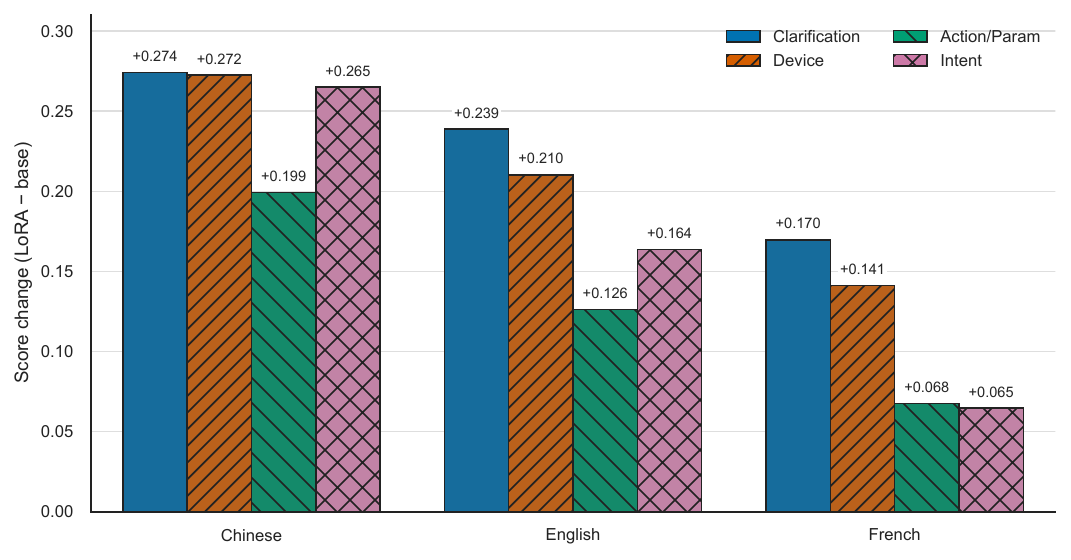}
    \caption{Language-specific performance of Qwen3.5-4B + LoRA on key grounding and intent metrics.}
    \label{fig:qwen-language}
\end{figure}

The held-out test set contained Chinese, English, and French samples, whereas the fine-tuning set contained only Chinese instances. The language-specific results therefore tested whether schema adaptation preserved cross-lingual generalization. Figure~\ref{fig:qwen-language} reports the principal grounding and intent metrics for Qwen3.5-4B + LoRA.

Qwen3.5-4B + LoRA improved across all three languages, with the largest gain in Chinese, the language used for fine-tuning. English and French also improved despite their absence from the fine-tuning set. This pattern provides evidence of cross-lingual transfer in schema following and household state grounding. However, the remaining cross-lingual performance gap leaves scope for multilingual fine-tuning. 

\begin{figure*}[ht!]
    \centering
    \includegraphics[width=0.95\textwidth]{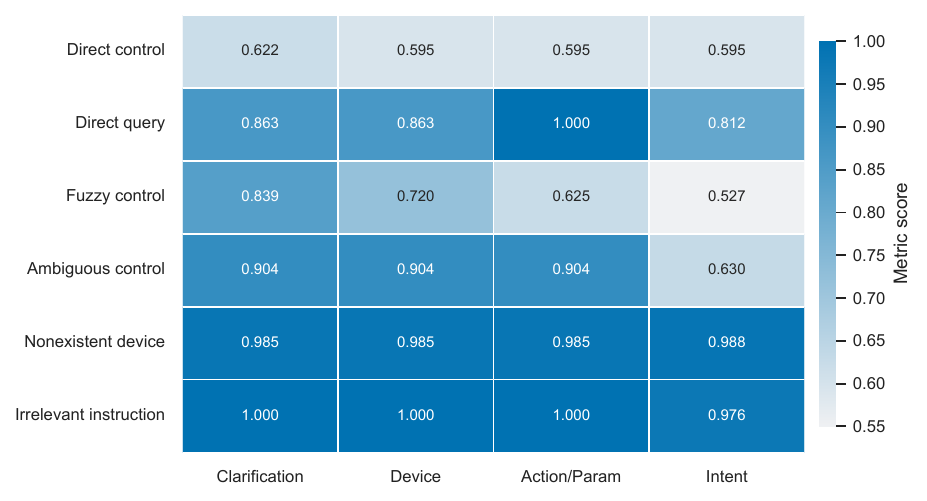}
    \caption{Scenario-level performance of Qwen3.5-4B + LoRA. Darker cells indicate stronger performance under each scenario metric pair.}
    \label{fig:qwen-scenario}
\end{figure*}

Figure~\ref{fig:qwen-scenario} reports the scenario-level performance of Qwen3.5-4B + LoRA. It focuses on clarification handling, device identification, action-parameter correctness, and instruction understanding. Together, these metrics measured whether the model generated the required SmartIntent action or, for non-executable cases, followed the expected clarification or execution-withholding behavior.

The model performed best on non-executable cases, including nonexistent device and irrelevant instruction scenarios, where it almost always withheld invalid commands. It also handled ambiguous control cases reliably. Errors were concentrated in executable cases, especially direct and fuzzy control. These scenarios required joint grounding of device, action, and parameters under household-state constraints. These lower action-parameter and instruction-understanding scores identify executable grounding as the main residual error after fine-tuning. 

Overall, Qwen3.5-4B + LoRA was the strongest compact model and approached the large local baseline on several schema-sensitive and execution oriented metrics. Gemma-4-E2B-it + LoRA did not reproduce these gains. Thus, the measured benefit of LoRA adaptation depended on the selected base model and training configuration. 
\subsection{Local deployment evaluation protocol}
\label{subsec:deployment-evaluation-protocol}

The model-level evaluation in Section~\ref{subsec:model-eval-results} establishes whether each model could satisfy the SmartIntent interface. We next evaluated the operational cost of the compact models in the local deployment described in Section~\ref{subsec:deployment-and-implementation}. The analysis addressed two questions. First, which non-thinking/instruct model provided the most reliable structured output? Second, did thinking mode improve output validity enough to justify its computational cost?

\subsubsection{Workloads, configurations, and repetitions}

Each experimental run processed one 50-request Chinese workload through a deployed OpenAI-compatible intent service. Requests were issued sequentially because every service used a single inference slot. The \texttt{default} and \texttt{stress} workloads had the same distribution over the six instruction categories shown in Fig.~\ref{fig:instruction-distribution}. The stress workload increased the input length and the amount of device state context, but it did not change the number of requests. We therefore use it to evaluate sensitivity to context load rather than to concurrent demand.

The non-thinking comparison included Qwen3.5-4B, the proposed Qwen3.5-4B + LoRA model, Gemma-4-E2B-it, Phi-4-mini, and Nemotron-4B. We compared thinking and non-thinking configurations separately for Qwen3.5-4B, the proposed model, and Gemma-4-E2B-it. Each model workload configuration was repeated between three and eleven times, depending on the available traces. Nemotron-4B is reported only for the default workload because the stress prompts exceeded its native 4,096-token context window, preventing us from obtaining a comparable stress workload trace. Individual runs are shown in the figures, and the error bars denote 95\% confidence intervals over repeated runs.

\subsubsection{Service-level metrics and statistical treatment}

We report five complementary service-level metrics. Valid outputs were parseable JSON objects that conformed to the SmartIntent schema and referenced only available devices, supported actions, and permitted parameters. The valid-output rate therefore measures execution interface validity, not complete semantic equivalence to the reference intent.

Mean response time measures end-to-end request latency. Request throughput is the number of completed requests divided by run duration. Completion throughput is the number of generated completion tokens divided by run duration. This distinction matters for thinking models, which can generate tokens rapidly while still increasing latency and reducing request throughput.

We first calculated each metric for one 50-request run and then averaged it across repeated runs. We calculated 95\% confidence intervals using a two-sided Student’s \textit{t} interval. Because repetition counts differed across configurations, comparisons should consider the individual observations, sample counts, and interval widths together.

\subsubsection{Idle power calibration and run-level energy}

\begin{figure}[t]
    \centering
    \includegraphics[width=0.78\linewidth]{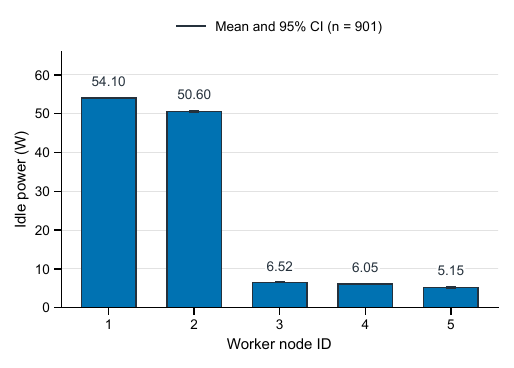}
    \caption{Idle power of the five worker nodes. Bars show the mean over 901 samples, and error bars show 95\% confidence intervals. Worker identifiers follow Table~\ref{tab:cluster}.}
    \label{fig:static-power}
\end{figure}

We measured idle power after deploying the experimental platform and before evaluating model inference. The five worker nodes had distinct idle power profiles as shown in Fig.~\ref{fig:static-power}. The two x86 GPU workers averaged 54.10~W and 50.60~W; the Jetson and Apple M4 workers ranged from 5.15~W to 6.52~W. 

Each estimate used 901 power samples. The corresponding 95\% confidence interval margins were at most 0.16\% of the mean for four workers and 2.77\% for the Apple M4 worker. These measurements provided stable node-specific baselines for estimating dynamic energy for each run $r$.

For run $r$, baseline-adjusted dynamic energy was calculated over the monitored worker nodes as
\begin{equation}
E_r = \frac{1}{3600}\sum_{t \in T_r}\sum_{n \in N_r}
\left(P_{r,n,t}-P^{\mathrm{idle}}_n\right)\Delta t,
\label{eq:run-energy}
\end{equation}
where $P_{r,n,t}$ is measured power, $P^{\mathrm{idle}}_n$ is the node-specific idle baseline, and $\Delta t=2$~s is the sampling interval. We retained negative baseline-adjusted samples to avoid upward bias from measurement noise. The reported energy values are therefore watt-hours per complete 50-request run. 
\subsection{Deployment performance in non-thinking mode}
\label{subsec:deployment-nonthinking}

\begin{figure*}[t]
    \centering
    \begin{minipage}[t]{0.49\textwidth}
        \centering
        \includegraphics[width=\linewidth]{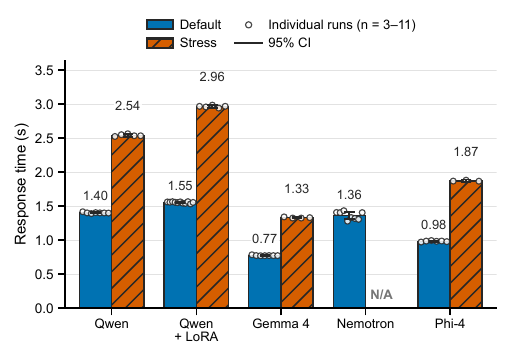}
        \small (a) Response time
    \end{minipage}
    \hfill
    \begin{minipage}[t]{0.49\textwidth}
        \centering
        \includegraphics[width=\linewidth]{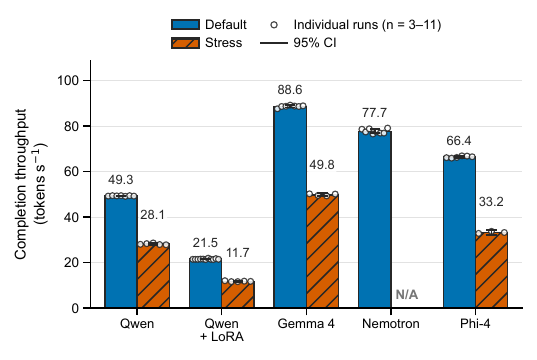}
        \small (b) Completion throughput
    \end{minipage}

    \begin{minipage}[t]{0.49\textwidth}
        \centering
        \includegraphics[width=\linewidth]{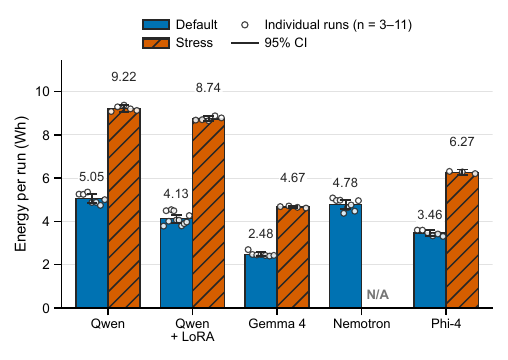}
        \small (c) Run-level energy
    \end{minipage}
    \hfill
    \begin{minipage}[t]{0.49\textwidth}
        \centering
        \includegraphics[width=\linewidth]{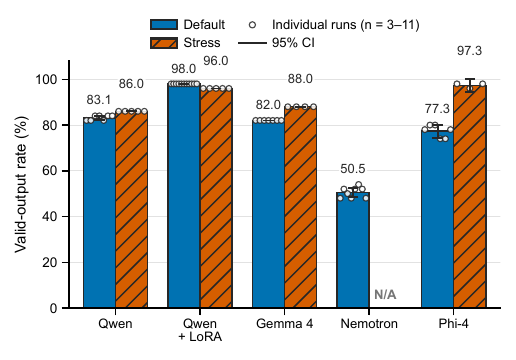}
        \small (d) Valid-output rate
    \end{minipage}
    \caption{Deployment-level comparison of non-thinking models on the default and stress workloads. Bars show means, circles show individual runs, and error bars show 95\% confidence intervals. Nemotron-4B has no stress workload result due to context limits.}
    \label{fig:nonthinking-deployment}
\end{figure*}

The non-thinking comparison reveals a clear reliability efficiency trade-off (Fig.~\ref{fig:nonthinking-deployment}). The proposed model achieved the highest and most consistent valid-output rate across the two workloads: 98.0\% on the default workload and 96.0\% on the stress workload. Its mean response time was 1.553~s and 2.965~s, respectively, corresponding to request throughputs of 0.644 and 0.337 requests~s$^{-1}$. Baseline-adjusted energy increased from 4.126~Wh per default run to 8.740~Wh per stress run, equivalent to 0.0825 and 0.1748~Wh per request.

Relative to its Qwen3.5-4B base model, the proposed model improved the valid-output rate by 14.9 percentage points on the default workload and 10.0 percentage points on the stress workload. This gain required a modest latency increase of 10.9\% and 16.6\%, respectively. Despite the higher latency, the proposed model consumed 18.4\% less run-level energy on the default workload and 5.2\% less on the stress workload. The fine-tuned model therefore improved interface reliability without introducing an energy penalty relative to its direct base model comparator.

Gemma-4-E2B-it was the fastest and least energy-intensive non-thinking service. It completed the default and stress requests in 0.773~s and 1.330~s on average and consumed 2.479~Wh and 4.667~Wh per run. These efficiency gains were accompanied by lower valid-output rates of 82.0\% and 88.0\%. Phi-4-mini reached 97.3\% validity under the stress workload, but only 77.3\% under the default workload, indicating less consistent interface behavior across the two test sets. Qwen3.5-4B fell between these extremes; Nemotron-4B produced only 50.5\% valid outputs on the default workload.

The completion throughput results should be interpreted separately from request capacity. The proposed model processed 21.47 completion tokens~s$^{-1}$ on the default workload and 11.74 completion tokens~s$^{-1}$ on the stress workload, lower than the other non-thinking models. Because this metric depends on both the number of generated tokens and total run time, it does not by itself show how many user requests the service can complete. In this single-slot sequential configuration, mean response time and request throughput measure service responsiveness more directly. 

Overall, the proposed non-thinking model provides the strongest operating point when execution interface validity is prioritized over minimum latency. Gemma-4-E2B-it is preferable when response time and energy dominate and a lower valid-output rate is acceptable. This distinction is important for SmartIntent because an invalid structured response is rejected before device dispatch; fast generation does not compensate for an unusable command payload. 
\subsection{Effects of context load and thinking mode}
\label{subsec:deployment-reasoning-stress}
Having established the non-thinking baseline, we next examine how context load and thinking mode affect output validity, latency, and energy. 

\subsubsection{Sensitivity to longer device context}

Increasing the input and device state context consistently raised operational cost. For the proposed non-thinking model, the stress workload increased mean response time by 90.9\%, from 1.553~s to 2.965~s, and increased run-level energy by 111.8\%, from 4.126~Wh to 8.740~Wh. Completion throughput fell by 45.3\%, from 21.47 to 11.74 tokens~s$^{-1}$. The valid-output rate changed by only two percentage points, from 98.0\% to 96.0\%. Thus, the main effect of the longer context on the proposed model was computational rather than structural: responses remained valid in most cases, but required substantially more time and energy.

The same operational pattern appeared across the other non-thinking services. From the default to the stress workload, response time increased by 81.5\% for Qwen3.5-4B, 72.1\% for Gemma-4-E2B-it, and 90.5\% for Phi-4-mini. Their run-level energy increased by 82.3\%, 88.3\%, and 81.4\%, respectively. The valid-output rate did not decrease uniformly and, for some models, increased on the stress set. Because the two workloads match category distributions but contain different cases, this variation should not be interpreted as evidence that longer contexts improve correctness. The robust conclusion is limited to operational cost: more context increased latency and energy for every non-thinking model with measurements under both workloads.

\subsubsection{Cost and benefit of thinking mode}

\begin{figure*}[t]
    \centering
    \begin{minipage}[t]{0.49\textwidth}
        \centering
        \includegraphics[width=\linewidth]{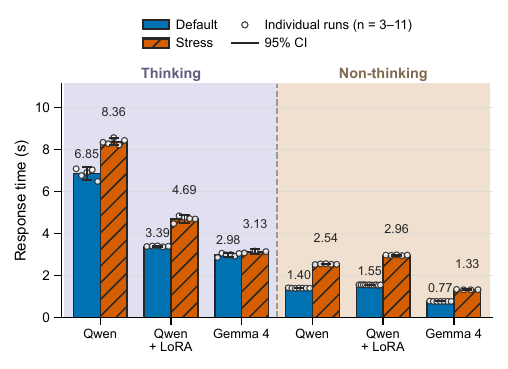}
        \small (a) Response time
    \end{minipage}
    \hfill
    \begin{minipage}[t]{0.49\textwidth}
        \centering
        \includegraphics[width=\linewidth]{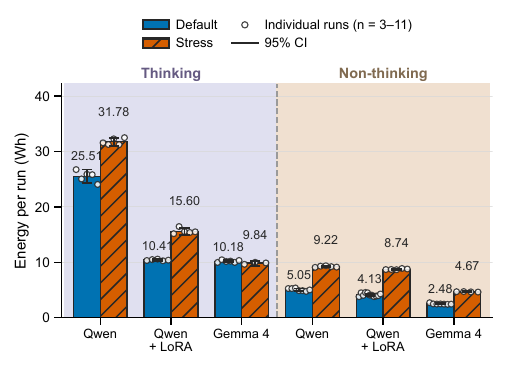}
        \small (b) Baseline-adjusted energy
    \end{minipage}

    \vspace{0.5em}
    \begin{minipage}[t]{0.49\textwidth}
        \centering
        \includegraphics[width=\linewidth]{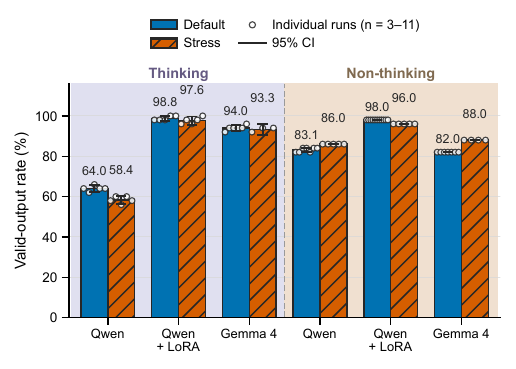}
        \small (c) Valid-output rate
    \end{minipage}
    \caption{Effect of thinking mode on response time, energy per 50-request run, and valid-output rate. Bars show means, circles show runs, and error bars show 95\% confidence intervals.}
    \label{fig:thinking-deployment}
\end{figure*}

Thinking mode did not provide a uniform deployment benefit (Fig.~\ref{fig:thinking-deployment}). For the proposed model, thinking increased valid-output rate by only 0.8 percentage points on the default workload and 1.6 percentage points on the stress workload. In exchange, response time increased by 118.2\% and 58.2\%, while energy increased by 152.2\% and 78.5\%, respectively. The proposed model already produced highly valid structured output without thinking, leaving little headroom for the additional reasoning tokens to improve this interface-level metric.

The effect depended strongly on model family. Enabling thinking for Gemma-4-E2B-it increased valid-output rate by 12.0 percentage points on the default workload and 5.3 percentage points on the stress workload, but increased latency by 285.1\% and 135.1\% and energy by 310.7\% and 110.7\%. For Qwen3.5-4B, thinking reduced valid-output rate by 19.1 percentage points on the default workload and 27.6 percentage points on the stress workload, while also increasing latency and energy. Thinking therefore cannot be assumed to improve a constrained JSON intent task. It must be evaluated against the exact serving template and output contract.

These results support non-thinking inference as the default SmartIntent deployment configuration. For the proposed model, it retains 96--98\% valid outputs while avoiding most of the latency and energy overhead associated with thinking. This conclusion is bounded to sequential, single-slot inference on the two 50-request workloads. The experiments do not measure cold-start latency, concurrent-request scaling, Knative autoscaling behavior, or end-to-end device actuation time. In addition, repetition counts vary across configurations, and Nemotron-4B lacks a stress trace. These factors limit claims about production-scale service capacity, but they do not change the observed within-configuration trade-off between output validity and operational cost.

\section{Conclusion}
\label{sec:Con}

This paper presented SmartIntent, a serverless edge intelligence architecture for local LLM-enabled building automation. SmartIntent frames natural language building control as an execution-oriented intent-to-execution problem, integrating local LLM intent services, device state context, a structured JSON interface, runtime validation, dispatching, rule management, and device control microservices. This design grounds user requests in available device context, represents them as machine-readable candidate actions, and validates them before downstream dispatch and, where applicable, rule management. The evaluation showed that compact local LLMs can support multilingual and ambiguity-aware intent interpretation when constrained by structured outputs and execution-oriented validation. The proposed Qwen3.5-4B+LoRA model achieved high output-format compliance, clarification handling, device identification, action-parameter correctness, and language consistency, narrowing the gap to larger local models on several execution-oriented metrics. In the local K3s/Knative deployment, it reached 98.0\% and 96.0\% valid-output rates on default and context stress workloads, with mean response times of 1.553~s and 2.965~s and dynamic energy of 4.126~Wh and 8.740~Wh per 50-request run. The results indicate that local LLM microservices with structured-output checks are a feasible direction for executable building automation under controlled edge deployment conditions, but they also expose deployment trade-offs. In the evaluated configurations, larger device state context and thinking mode increased latency and energy cost, so local LLM automation should be evaluated not only by intent accuracy but also by runtime and energy behavior. Future work will study concurrent and autoscaling deployments, real building integration, broader multilingual adaptation, and more complete validation of persistent automation rules. 

\section*{Declarations}

\bmhead{Funding:}
No funding was received for this study.



\footnotesize
\bibliography{BibTeX}

\end{document}